\documentclass[paper]{JHEP3}
\pdfoutput=1
\usepackage{amsmath,amssymb,amsthm,amscd,graphicx}
\input epsf.sty

\newtheorem{theorem}{Theorem}[section]

\theoremstyle{definition}

\newtheorem{remark}[theorem]{Remark}

\newcommand{\CB}{{\cal B}}
\newcommand{\CC}{{\cal C}}

\newcommand{\CF}{{\cal F}}
\newcommand{\CG}{{\cal G}}
\newcommand{\CH}{{\cal H}}
\newcommand{\CI}{{\cal I}}

\newcommand{\CN}{{\cal N}}
\newcommand{\CO}{{\cal O}}
\newcommand{\CP}{{\cal P}}

\newcommand{\CR}{{\cal R}}
\newcommand{\CS}{{\cal S}}
\newcommand{\CT}{{\cal T}}

\def\IP{{\mathbb P}}

\def\IS{{\mathbb S}}

\newcommand{\tr}{{\rm Tr}}
\newcommand{\re}{{\rm e}}
\newcommand{\ri}{{\rm i}}
\newcommand{\rd}{{\rm d}}

\newcommand{\Tr}{\mathop{\rm Tr}\nolimits}
\newcommand{\Li}{\mathop{\rm Li}\nolimits}

\newcommand{\wigner}{{\mathrm{W}}}

\newcommand{\be}{\begin{equation}}
\newcommand{\ee}{\end{equation}}
\newcommand{\ba}{\begin{aligned}}
\newcommand{\ea}{\end{aligned}}
\newcommand{\ben}{\begin{eqnarray}\displaystyle}
\newcommand{\een}{\end{eqnarray}}

\newcommand{\sectiono}[1]{\section{#1}\setcounter{equation}{0}}

\newdimen\tableauside\tableauside=1.0ex
\newdimen\tableaurule\tableaurule=0.4pt
\newdimen\tableaustep
\def\phantomhrule#1{\hbox{\vbox to0pt{\hrule height\tableaurule width#1\vss}}}
\def\phantomvrule#1{\vbox{\hbox to0pt{\vrule width\tableaurule height#1\hss}}}
\def\sqr{\vbox{%
  \phantomhrule\tableaustep
  \hbox{\phantomvrule\tableaustep\kern\tableaustep\phantomvrule\tableaustep}%
  \hbox{\vbox{\phantomhrule\tableauside}\kern-\tableaurule}}}
\def\squares#1{\hbox{\count0=#1\noindent\loop\sqr
  \advance\count0 by-1 \ifnum\count0>0\repeat}}
\def\tableau#1{\vcenter{\offinterlineskip
  \tableaustep=\tableauside\advance\tableaustep by-\tableaurule
  \kern\normallineskip\hbox
    {\kern\normallineskip\vbox
      {\gettableau#1 0 }%
     \kern\normallineskip\kern\tableaurule}%
  \kern\normallineskip\kern\tableaurule}}
\def\gettableau#1{\ifnum#1=0\let\next=\null\else
\squares{#1}\let\next=\gettableau\fi\next}

\newcommand{\figref}[1]{Fig.~\protect\ref{#1}}

\title{ABJM Wilson loops in the Fermi gas approach}

\author{Albrecht Klemm$^1$, Marcos Mari\~no$^2$, Marc Schiereck$^1$ and Masoud\ Soroush$^1$
\\
$^1$ Bethe Center for Theoretical Physics, Physikalisches Institut\\
der Universit\"at Bonn, Nussallee 12, D-53315 Bonn, Germany\\
\\
$^2$ D\'epartement de Physique Th\'eorique et Section de Math\'ematiques,\\
Universit\'e de Gen\`eve, Gen\`eve, CH-1211 Switzerland\\
\\
\email{aklemm@th.physik.uni-bonn.de},\, \email{marcos.marino@unige.ch}\\
\email{marc@th.physik.uni-bonn.de},\, \email{soroush@uni-bonn.de}\\
}

\preprint{Bonn-TH-12-12}

\abstract{The matrix model of ABJM theory can be formulated in terms of an ideal Fermi gas with a non-trivial one-particle Hamiltonian. We show that, in this formalism, vevs
of Wilson loops correspond to averages of operators in the statistical-mechanical problem. This makes it possible
to calculate these vevs at all orders in $1/N$, up to exponentially small corrections, and for arbitrary Chern--Simons coupling, by using the
WKB expansion. We present explicit results for the vevs of 1/6 and the 1/2 BPS
Wilson loops, at any winding number, in terms of Airy functions. Our expressions are shown to reproduce the low genus results obtained previously in the 't Hooft expansion.
}

\begin{document}

\sectiono{Introduction}

Localization techniques in superconformal field theories have provided matrix model representations for partition functions and Wilson loop vacuum expectation values (vevs)
on spheres. For $\CN=4$ super Yang--Mills theories, these techniques were developed in \cite{pestun}, providing a proof of previous conjectures in \cite{esz,dg} which
proposed a Gaussian matrix model formula for the vev of a $1/2$ BPS, circular Wilson loop. This was extended to Chern--Simons--matter theories in \cite{kwy,jafferis,hama}.
In particular, a matrix model was obtained in \cite{kwy} which calculates the partition function and the vev of the $1/6$ BPS Wilson loops for ABJM theory \cite{abjm}
constructed in \cite{dp,cw,rey}. $1/2$ BPS Wilson loops were constructed and localized in \cite{dp}, and their vevs are calculated by computing the averages of supertraces in the ABJM matrix model of \cite{kwy}.

Once the matrix models have been written down, an important question is to extract from them
the $1/N$ expansion of the observables, in order to test predictions based on the AdS/CFT correspondence.
In the case of the $1/2$ BPS Wilson loop of $\CN=4$ super Yang--Mills, this is relatively straightforward, since the matrix model is a Gaussian one. In particular, in \cite{dg}, a procedure was presented to obtain the full $1/N$ expansion of the $1/2$ BPS Wilson loop, and explicit expressions were obtained for the leading term in the 't Hooft parameter, at all orders in $1/N$. This term gives, in the AdS dual, the leading contribution coming from strings with one boundary and arbitrary genus.

The ABJM matrix model is much more complicated than the Gaussian matrix model. However, its free energy can be computed to any desired order in the 't Hooft $1/N$ expansion
\cite{dmp,dmpnp}, in a recursive way. This is achieved by using the holomorphic anomaly equations of topological string theory \cite{bcov}, as adapted
to matrix models and local geometries in for example \cite{hk,emo,hkr}.
For Wilson loops, results at low genus can also be obtained from matrix model techniques. The exact planar result was obtained in \cite{mpwilson}, and the first $1/N$ correction was calculated in \cite{dmp} by using the results of \cite{akemann}. In principle, one can compute higher genus corrections by using for example the topological recursion of \cite{eo}, but this procedure becomes rapidly quite cumbersome. Unfortunately, we lack an efficient holomorphic anomaly equation for open string amplitudes which makes possible to go beyond the very first genera.

In the context of ABJM theory, understanding the full $1/N$ expansion is however of great interest, since this gives quantitative information about the M-theory AdS dual. Equivalently, one can
try to compute the observables in the so-called {\it M-theory expansion}. In this expansion, one still considers the limit of large $N$ but $k$ (the Chern--Simons coupling, or equivalently the inverse string coupling constant) is fixed. In \cite{fhm}, building on the results of \cite{dmp,dmpnp}, it was shown that the full $1/N$ expansion of the partition function could be summed up into an Airy function, after neglecting exponentially small corrections. This raises the question of finding a method for analyzing the matrix model directly in the M-theory regime, without having to resum the 't Hooft expansion. The method developed in \cite{hkpt} works directly in the M-theory regime and can be applied to a large class of Chern--Simons--matter theories, but in its current form it is only valid in the strict large $N$ limit.

A systematic method to analyze the matrix models arising in $\CN\ge 3$ Chern--Simons--matter theories, in the M-theory expansion, was introduced in \cite{mp}. The basic idea of the method is to reformulate the matrix model partition function, as the partition function of a non-interacting, one-dimensional Fermi gas of $N$ particles, but with a non-trivial quantum Hamiltonian. In this reformulation, the Chern--Simons coupling $k$ becomes the Planck constant $\hbar$, and the M-theory expansion corresponds to the thermodynamic limit of the quantum gas. It was shown in \cite{mp} that the partition function of the gas could be computed, at all orders in $1/N$, by doing the WKB approximation to next-to-leading order (neglecting exponentially small corrections). This makes it possible to re-derive the Airy function behavior found in \cite{fhm} for ABJM theory, and generalize it to a large class of $\CN=3$ Chern--Simons--matter theories. The
Fermi gas approach provides as well an elementary and
physically appealing explanation of the famous $N^{3/2}$ scaling predicted in \cite{kt} and first proved in \cite{dmp}: it is the expected scaling for a Fermi gas with a linear dispersion relation and a linear
confining potential.

In this paper we extend the Fermi gas approach of \cite{mp} to the calculation of vevs of $1/6$ and $1/2$ BPS Wilson loops. As expected, the vevs correspond, in the statistical-mechanical formulation, to averages of $n$-body operators. Since the gas is non-interacting, this can be reduced to a quantum-mechanical computation in the one-body problem, which can be in principle done in the semiclassical expansion. However, in this case a precise determination of the vev requires the resummation of an infinite number of quantum corrections. This is not completely unexpected: already in the calculation of the partition function in \cite{mp}, there is an overall factor which is a non-trivial function of $k$ and involves a difficult, all-order calculation of quantum corrections. For Wilson loops in ABJM theory, one can actually perform the resummation directly, and obtain a closed formula for the $1/N$ expansion of the $1/6$ and $1/2$ BPS Wilson loops in terms of Airy functions. In the case of a $1/2$ BPS Wilson loop in the
fundamental representation, the result for the normalized vev is particularly nice,
\be
\label{wilson12}
\langle W_{\tableau{1}}^{1/2}\rangle=\frac{1}{4} \csc\left({2 \pi \over k}\right)  \frac{\mbox{Ai}\Big[C^{-1/3} \Big(N-\frac{k}{24}
-\frac{7}{3 k }\Big)\Big]}{\mbox{Ai}\Big[C^{-1/3}\Big(N-\frac{k}{24}
-\frac{1}{3k }\Big)\Big]}
\ee
where
\be
C=\frac{2}{\pi^{2}k}.
\ee

This result is exact at all orders in $1/N$, up to exponentially small corrections in $N$ (corresponding to world sheet or membrane instanton
corrections). The denominator in (\ref{wilson12}) is the partition function of ABJM theory as computed in \cite{fhm,mp}.
The corresponding expression for the $1/6$ BPS Wilson loop, and for arbitrary winding, is more involved, and it is given below in section 4 (eq. (\ref{airy})).

The paper is organized as follows. In section 2, we start with a brief review of certain aspects of ABJM matrix model, and in particular, we review and extend the results of matrix model computations for the 1/6 and 1/2 BPS Wilson loop expectation values at genus zero and one. In section 3, we first proceed by recalling some standard techniques of quantum statistical mechanics in phase space, which are going to be used later on in this paper. We then turn into a brief review of the Fermi gas approach which was originally introduced in \cite{mp}. Section 4 is the core of our paper. We first demonstrate in 4.1 how we can include the Wilson loops in the Fermi gas formalism. We continue in section 4.2 by first computing the full quantum corrected Hamiltonian of the fermionic system, and then by calculating the corresponding Wigner-Kirkwood corrections for the quantum mechanical averages. In 4.3, we deal with the integration over the quantum corrected Fermi surface, and section 4.4 contains the explicit results for Wilson loop vevs and a detailed comparison with the 't~Hooft expansion in the strong coupling regime. Section 5 is devoted to conclusions and prospects for future work. In Appendix A, we present the details of the matrix model computation for the 1/6 BPS Wilson loop correlator at arbitrary winding. Appendix B summarizes the results of the 'tHooft expansion at genus three and genus four.

\sectiono{Wilson loops in ABJM theory}

\subsection{$1/6$ BPS and $1/2$ BPS Wilson loops}

The ABJM theory \cite{abjm,abjmreview} is a quiver Chern--Simons--matter theory in
three dimensions with gauge group $U(N)_k \times U(N)_{-k}$ and $\CN=6$ supersymmetry.
The Chern--Simons actions have couplings $k$ and $-k$, respectively, and the theory contains four bosonic fields $C_I$, $I=1, \cdots, 4$,
in the bifundamental representation of the gauge group. One can construct an extension of this theory \cite{abj}
with a more general gauge group $U(N_1)_k \times U(N_2)_{-k}$, but we will not consider it in detail in this paper. The 't Hooft parameter of this theory is
\be
\lambda={N\over k}.
\ee
A family of Wilson loops in this theory has been constructed in \cite{dp,cw,rey}, with the structure
\be
\label{wloop}
W^{1/6}_R = \tr_R \, \CP \, \exp \int \left( \ri A_\mu  \dot x^\mu + {2\pi \over k} |\dot x| M^I_J C_I \bar C^J \right) \rd s\ ,
\ee
where $A_\mu$ is the gauge connection in the $U(N)_k$ gauge group of the first node, $x^\mu(s)$
is the parametrization of the loop, and
$M_I^J$ is a matrix determined by supersymmetry. It can be chosen so that, if the geometry of the loop is a line or a circle, four real supercharges are preserved.
Therefore, we will call (\ref{wloop}) the 1/6 BPS Wilson loop. A similar construction exists for a loop based on the other gauge group, and one obtains
a Wilson loop associated to the second node
\be
\label{wloopnar}
\widehat W^{1/6}_R\ .
\ee

In \cite{kwy} it was shown, through a beautiful application of localization
techniques, that both the vev of (\ref{wloop}) and the partition function on the three-sphere can be computed by a matrix model (see \cite{lectures} for a pedagogical review). This matrix model is defined by the partition function
\be
\label{kapmm}
Z_{\rm ABJM}(N, g_s)={1\over N!^2} \int \prod_{i=1}^{N}{ \rd \mu_i \rd \nu_j  \over (2 \pi)^2} {\prod_{i<j} \sinh^2 \left( {\mu_i -\mu_j \over 2}\right)  \sinh^2 \left( {\nu_i -\nu_j \over 2}\right) \over
\prod_{i,j}  \cosh^2 \left( {\mu_i -\nu_j \over 2}\right)} \re^{-{1\over 2g_s}\left(  \sum_i \mu_i^2 -\sum_j \nu_j^2\right)},
\ee
where the coupling $g_s$ is related to the Chern--Simons coupling $k$ of ABJM theory as
\be
g_s={2 \pi \ri \over k}.
\ee
One of the main results of \cite{kwy} is that the normalized vev of the $1/6$ BPS Wilson loop (\ref{wloop}) is given by a normalized correlator in the matrix model (\ref{kapmm}),
\be
\label{16WL}
\langle W_R^{1/6}\rangle = \left\langle \tr_R\left (\re^{\mu_i}\right)  \right\rangle_{\rm ABJM},
\ee
Notice that the Wilson loop
for the other gauge group,
\be
\label{other}
\langle \widehat W^{1/6}_R\rangle =\left\langle \tr_R \left( \re^{\nu_i} \right) \right\rangle_{\rm ABJM}
\ee
can be obtained from (\ref{16WL}) simply by conjugation, or equivalently, by changing the sign of the coupling constant $g_s\rightarrow -g_s$. From now on we will then focus,
without loss of generality, on the Wilson loop associated to the first node, and we will also assume that $k>0$ in the first node.

The Wilson loop (\ref{wloop}) breaks the symmetry between the two gauge groups. A class of 1/2 BPS Wilson loops was constructed in
\cite{dt} which treats the two gauge groups in a
more symmetric way (see also \cite{wilsonk}). These loops have a natural supergroup structure in which the quiver gauge group $U(N) \times U(N)$ is promoted
to $U(N|N)$, and they can be defined in any super-representation $\CR$.
In \cite{dt} it has been argued that this 1/2 BPS loop, which we will denote by $W^{1/2}_\CR$, localizes to the matrix model correlator
\be
\label{12wl}
\langle W^{1/2}_\CR \rangle=\left\langle {\rm Str}_\CR \begin{pmatrix}\re^{\mu_i} &0 \\ 0& -\re^{\nu_j} \end{pmatrix} \right\rangle_{\rm ABJM}
\ee
in the ABJM matrix model. Here, $ {\rm Str}_\CR$ denotes a super-trace in the super-representation $\CR$. In order to write this in more down-to-earth terms, we note
that a representation of $U(2N)$ induces a super-representation of $U(N|N)$, defined by the
same Young tableau $\CR$ (see
for example \cite{bars}). Therefore, (\ref{12wl}) can be also written as \cite{bars}
\be
\label{super-ordi}
\ba
{\rm Str}_\CR\begin{pmatrix} \re^{\mu_i} & 0\\ 0& -\re^{\nu_j} \end{pmatrix}& =\sum_{\vec k} {\chi_\CR(\vec k) \over z_{\vec k}}  \prod_\ell
\left( {\rm Str}\begin{pmatrix} \re^{\mu_i} & 0\\ 0& -\re^{\nu_j} \end{pmatrix}^\ell \right)^{k_\ell}\\
&=\sum_{\vec k} {\chi_\CR(\vec k) \over z_{\vec k}}
\prod_\ell \left( \tr  \left( \re^{\ell \mu_i}\right) -(-1)^\ell \tr \left( \re^{\ell \nu_j} \right)
\right)^{k_\ell}.
\ea
\ee
In this equation, which is the supergroup generalization of Frobenius formula, $\vec k=(k_\ell)$ is a vector of non-negative, integer entries, which can be regarded as a conjugacy class of the symmetric group, $\chi_\CR(\vec k)$ is the character of this conjugacy class in the representation $\CR$, and
\be
z_{\vec k}=\prod_\ell \ell^{k_\ell} k_\ell!\, .
\ee
We will be particularly interested on Wilson loops with winding number $n$,  which in the basis of representations are defined by
\be
W_n^{1/6}  = \sum_{s=0}^{n-1} (-1)^s W_{R_{n,s}}^{1/6}.
\ee
Here, $R_{n,s}$ is a ``hook'' representation with $n$ boxes in total, $n-s$
boxes in the first row, and one box in the remaining rows. For $n=1$ we recover the usual Wilson loop in the fundamental
representation. In terms of matrix model vevs,
\be
\label{n-corr}
\langle W_n^{1/6}\rangle =\left\langle \tr\, \left (\re^{n\mu_i}\right)  \right\rangle_{\rm ABJM}.
\ee
In view of (\ref{super-ordi}), the $1/2$ BPS Wilson loop with winding $n$ is simply given by
\be
\label{sumw}
\langle W^{1/2}_n \rangle = \langle W^{1/6} _n\rangle-(-1)^n  \langle \widehat W^{1/6}_n\rangle.
\ee
In general, as it is clear from (\ref{super-ordi}), the vevs of 1/2 BPS Wilson loops can be obtained if we know the vevs of 1/6 BPS Wilson loops, but the former are much simpler.

\subsection{The geometry of the ABJM theory}

In~\cite{mpwilson,dmp} the ABJM partition function and the Wilson loop vevs are mapped, via the
spectral curve of the lens space matrix model, to geometric
invariants of the elliptic curve
\be
H(X,Y)=X+\frac{1}{\varphi_1^2 X } +Y+\frac{1}{\varphi_2^2 Y}+1=0,
\label{mirrorcurvep1xp1}
\ee
which are in turn related to meromorphic differentials of the third kind, see \cite{lectures} for a review. 
In particular, in the planar limit, the partition function and the Wilson loop vevs are related to periods of
these differentials. The higher $N$ corrections are related to these periods by
a recursive procedure, which amounts to integration of the loop
equations of the matrix model \cite{eynard,eo}. In (\ref{mirrorcurvep1xp1}) $X,Y$ are
$\mathbb{C}^*$ variables and eqn. (\ref{mirrorcurvep1xp1}) is the B-model
mirror  curve of the local Calabi-Yau geometry
$M_{cy}={\cal O}(-K_{\mathbb{P}^1\times \mathbb{P}^1})\rightarrow \mathbb{P}^1\times
\mathbb{P}^1$, i.e. the total space of the anti canonical line bundle over
$\mathbb{P}^1\times \mathbb{P}^1$.

After multiplying (\ref{mirrorcurvep1xp1}) with $XY$, homogenizing
it to a cubic with $W$, swaping $W$ with $-Y$ and rescaling
$X\mapsto X \varphi_1$, one gets the curve
\be
\tilde H(X,Y)=Y^2-Y\left(1+X\varphi_1+X^2 \frac{\varphi^2_1}{\varphi^2_2}\right)+X^2 =0
\label{mirrorcurvep1xp1b}
\ee
One might parameterize the $\mathbb{C}^*$ variables $X=\re^u$ and $Y=\re^v$.
Then the relevant meromorphic differentials of the third kind are given by
\be
\mu_k= v \re^{k u} \rd u = \log(Y) X^{k-1} \rd X, \qquad \quad k=0,1,\ldots ,
\label{meromorphicdiffsonc*}
\ee
where
\be
Y= \frac{a(X)}{2} \pm \frac{1}{2} \sqrt{\sigma(X)}\ .
\label{Y}
\ee
This form is typical of local mirror geometries. With the above parameterization the discriminant
is given as
\be
\sigma(X)=\prod_{i=1}^4 (X-x_i)=a(X)^2- 4 X^2,\qquad {\rm with} \quad a(X)=1+X\varphi_1+X^2\frac{\varphi^2_1}{\varphi^2_2}\ .
\ee
The branch points involve square roots of the $\varphi_i$, but with an
appropriate ordering one has
\be
\varphi_1=-\frac{1}{2}\sum_{i=1}^4 x_i,\qquad
\frac{\varphi_1}{\varphi_2}=\frac{1}{4}(x_1+x_2-x_3-x_4),
\qquad x_1=\frac{1}{x_2}=:a, \qquad x_4=\frac{1}{x_3}=:-b \ .
\ee

Note that (\ref{mirrorcurvep1xp1b},\ref{Y}) defines the same
family of (hyper) elliptic curves as
\be
y^2=\sigma(x)\ ,
\label{hyperelliptic}
\ee
where we identified $X,Y$ with $x,y$. This identification amounts
to a compactification of the $\mathbb{C}^*$ variables $X,Y$ and does
not affect integrals over closed cycles, up to one important
subtlety: at $X\rightarrow \infty$, $\mu_0$ behaves like
\be
\mu_0(X)=\frac{2}{X}(\log\left(\frac{\varphi_1}{\varphi_2}\right)+\log(X))
+\frac{1}{X^2}\frac{\varphi_2^2}{\varphi_1}-
\frac{1}{X^3}\left(\frac{\varphi_2^4}{\varphi_1^4}-\frac{\varphi_2^4}{2
    \varphi_1^2}+\frac{\varphi_2^2}{\varphi_1^2}\right)+  {\cal
  O}\left(\frac{1}{X^4}\right)\ .
\label{residuaatinfinity}
\ee
In the compactification one has to regularize the form $\mu_0$ to
\be
\mu_0(x)=\mu_0(X)|_{x=X}-\frac{2}{x}\log(x).
\ee
Derivatives of $\mu_0(x)$ w.r.t. to $\varphi_i$ are related to standard elliptic
integrals on (\ref{hyperelliptic}).

When the ranks of the nodes in ABJM theory are not identical (this is the so-called ABJ theory \cite{abj}), there are two 't Hooft parameters defined by 
\be
\lambda_i=\frac{N_i}{k}, \qquad i=1,2. 
\ee
In the Calabi--Yau picture, these parameters 
are mirror coordinates and as such they are identified
with the periods
\be
\lambda_i=\frac{1}{4 \pi \ri} \int_{\CC_i} \mu_0,  
\label{periods}
\ee
where the cycles have the geometry
\be
\label{Zend}
\CC_1=(1/a, a), \qquad \CC_2=(-b, -1/b) \ .
\ee

\begin{figure}
\center
\includegraphics[height=4cm]{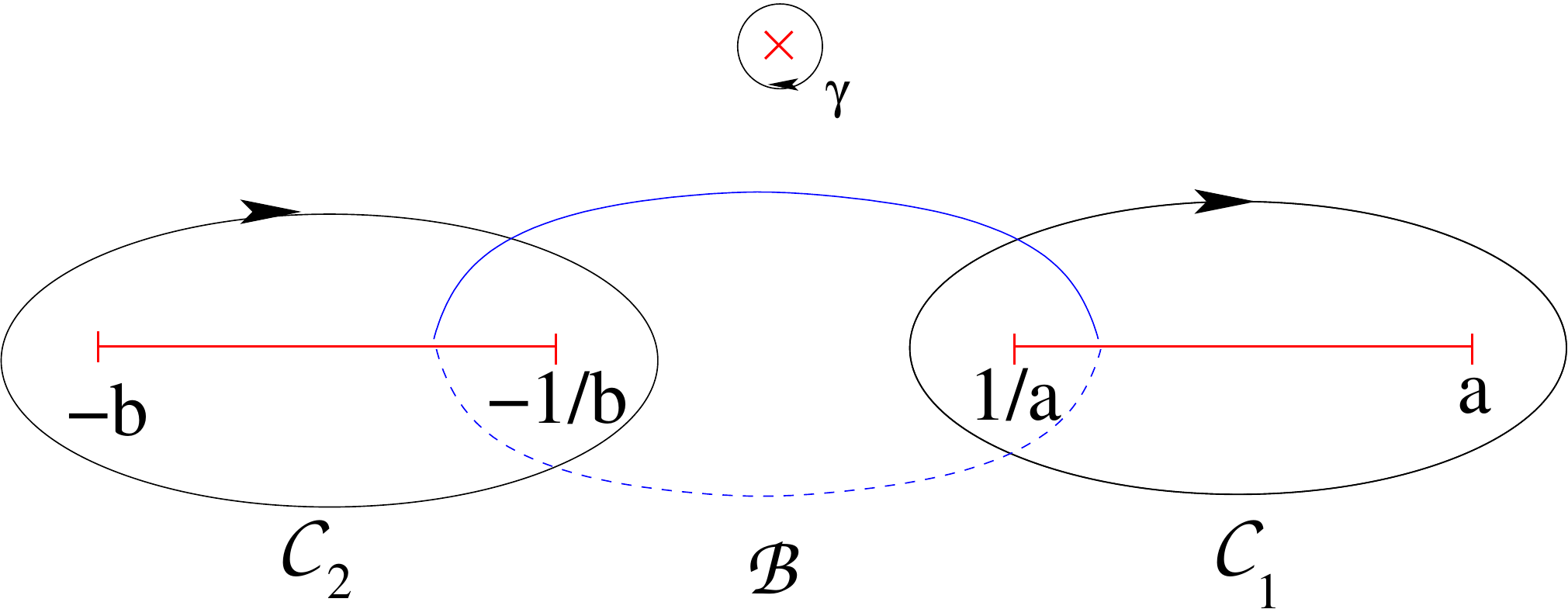}
\caption{The cycles in the ABJM geometry in the $x$-plane.
The non-vanishing residua of the forms at $x=\infty$ }
\label{cycles}
\end{figure}

The homology relations imply that the $\CC_i$ periods
are non identical because of the pole in the $\mu_k$.
In particular for $\mu_0$ it is clear from \figref{cycles} 
and (\ref{residuaatinfinity}) that there is an exact relation between
the periods (\ref{periods})
\be
\exp(2 \pi \ri (\lambda_1-\lambda_2))=\frac{\varphi_1}{\varphi_2}\ .
\ee
For this reason, the ABJM slice 
\be
\lambda_1=\lambda_2\ {\rm mod} \ \mathbb{Z}
\ee
can be identified with an algebraic submanifold of the 
complex deformation space of (\ref{mirrorcurvep1xp1}). This submanifold is simply given by
\be
\varphi_1=\varphi_2=\varphi=\ri \kappa.
\ee
In particular, in the slice one has
\be
\partial_\varphi \mu_k=\omega_k= \frac{x^k}{\sqrt{\sigma(x)}} \rd x\ ,
\label{relationbetweendifferentials}
\ee
i.e. all closed integrals of $\mu_k$ on (\ref{mirrorcurvep1xp1b},\ref{Y})
are determined up to a constant by standard elliptic integrals on
(\ref{hyperelliptic}).
For latter reference we note that the parameterization
of the branch points by $\kappa$ is
 \be
  a(\kappa)=\frac{1}{2} \left(2+ \ri \kappa+\sqrt{\kappa(4 \ri -\kappa)}
  \right),\qquad
  b(\kappa)=\frac{1}{2} \left(2-\ri \kappa+\sqrt{-\kappa(4\ri +\kappa)}
  \right)\ .
  \ee

On the slice (\ref{mirrorcurvep1xp1b}) is
an algebraic family  of elliptic curves with monodromy group $\Gamma^0(4)$ and 
$j$-invariant
\be
j=\frac{16-16 \varphi^2+\varphi^4}{1728 \varphi^2(16-\varphi^2)}\ .
\ee
This family is related to the $\Gamma_0(2)$ curve of pure
$SU(2)$ SW-theory 
\be
y^2=(x^2-u)^2-\Lambda^4
\ee
by identifying 
\be
u =\pm \left(1 - \frac{\varphi^2}{8} \right)\Lambda^2.
\ee
Indeed, the period integrals of $\mu_0$ are annihilated by a single
Picard--Fuchs differential operator for
$M_{cy}$, after identifying the K\"ahler classes of the
$\mathbb{P}^1$ i.e. $T_1=T_2$ (in the notation of \cite{dmp}). 
It reads\footnote{The formulas
$\theta_x=a\theta_y$ if $y=x^a$ and $[\theta_x,x^z]=a x^a$ make
the comparison with \cite{kz} trivial.}
\be
{\cal D}= (\varphi^2 \theta_{\varphi}^2- 16 (2
\theta_{\varphi}-1)^2)\theta_{\varphi}=
\varphi (\varphi^2(
\theta_{\varphi}+1)^2+16\theta_{\varphi}^2)\partial_\varphi=
\varphi {\cal D}_{hol} \partial_\varphi ,
\label{Pfequation}
\ee
where $\theta_x=x \rd/\rd x$ is the logarithmic derivative.
${\cal D}_{hol}$ annihilates the periods over the holomorphic
differential $\omega_0$ on (\ref{hyperelliptic}), as a
consequence of (\ref{relationbetweendifferentials}). The differential
equation (\ref{Pfequation}) has three critical points: $\varphi^2=0$,
$\varphi^2=16$ and $\varphi^2=\infty$. Let us describe the behavior
of the periods at these points and determine the  analytic
continuations and the monodromy action. The weak coupling point
of ABJM is the point $\varphi=0$. In the $w=\varphi^2$ variable
the period basis looks like
\be
\Pi= \left(\begin{array}{c} \int_\gamma \mu_0\\ \int_{\cal B} \mu_0\\
\int_{\cal C} \mu_0  \end{array} \right)= \left(\begin{array}{c} 1 \\ \partial_\lambda F^0   \\ \lambda \end{array}
\right),\ \  {\rm with}\ \
\left(\begin{array}{c} 1\\ \lambda \\ \tilde F_{\lambda}   \end{array} \right)
= \left(\begin{array}{c} 1\\ \frac{\sqrt{w}}{8 \pi \ri}[1 + \frac{w}{192} +{\cal O}(w^2)]\\
\frac{2\lambda}{\pi \ri} \log(w) + \frac{\sqrt{w}}{4 \pi^2 \ri} [\frac{5}{588} w + {\cal O}(w^2)]
  \end{array}\right)\ .
\ee
where 
\be
\partial_\lambda F^0= \tilde F_{\lambda}-(2+\ri b)\lambda
-\frac{1}{2}, \qquad b=\frac{4\log(2)+1}{\pi}.
\ee
The recursion defining $\lambda$ can be summed up to yield \cite{mp}
\be
 \label{lamkap}
 \lambda={\kappa \over 8 \pi}   {~}_3F_2\left(\frac{1}{2},\frac{1}{2},\frac{1}{2};1,\frac{3}{2};-\frac{\kappa^2
   }{16}\right)\ .
   \ee
This function plays the role of the mirror map at the orbifold, while
$\partial_\lambda F^0_w$ is the dual period. This pair defines the genus zero prepotential $F^0_w$, by
special geometry, as well as the polarization on the ABJM slice.

The point  $\varphi^2=\infty$ is the strong coupling point of
ABJM theory, $\lambda\rightarrow \infty$. It corresponds to the large radius
point of topological string theory. The topological string basis is obtained
by the local limit of a compact Calabi-Yau manifold, and it is half integral
in the homology of the curve (\ref{mirrorcurvep1xp1b})
\be
\Pi =\left(\begin{array}{ccc} 1& 0&0 \\ -1&1& 0\\
      0&-\frac{1}{2}&1\end{array}\right)\Pi_{ts}\ .
\ee
In the coordinates $z=\varphi^{-2}$ the topological string basis reads
\be
\Pi_{ts}=\left(\begin{array}{c} 1\\ T\\ \partial_T F_{gw}^0 \end{array}
\right)=\left(\begin{array}{c} 1\\
\frac{1}{2\pi \ri}[\log(z)+ 4z +{\cal O}(z^2)] \\
-\frac{1}{2} \left(\frac{1}{2\pi \ri}\right)^2[\log^2(z)+ 4z \log(z) +8 z +{\cal
  O}(z^2)] -\frac{1}{12}  \end{array}\right)\ .
\ee
Here, $Q=\exp(2 \pi \ri T)$, and $\partial_T F^0_{gw}$ can be integrated
to obtain the genus $0$ prepotential 
\be
F^0_{gw}(Q)=-\frac{1}{6} T^3 -\frac{1}{12} T +
c+\sum_{d=1}^\infty n^9_d {\Li}_3(q^d). 
\ee
This is the generating function
of $g=0$ BPS invariants, summing over the degrees $d_1+d_2=d$ w.r.t. to both K\"ahler classes
of the $\mathbb{P}^1$'s\footnote{Up to a constant $c=\frac{\chi}{2 (2 \pi \ri)^3}$,
which depends on the regularized Euler number of the local geometry for $\chi=4$.}.
Near the conifold point, and in the $u=(1-\frac{16}{\varphi^2})$ coordinates, the basis reads \footnote{The irrational
 constant $c=.3712268727\ldots$ is fastest iterated using the Meijers function \cite{kz}.}
\be
\Pi_{ts}=\left(\begin{array}{c} 1\\
\partial_{T_c} F_c^0\\
T_c \end{array} \right)=
\left(\begin{array}{c} 1\\ \frac{1}{2\pi^2 \ri}[4\pi T_c\log(u) + \frac{9}{16} u^2+{\cal O}(u^3)] +2 b \ri T_c+c   \\
\frac{1}{4\pi}[u+ \frac{5}{8}u^2  + {\cal O}(u^3)] \end{array}\right)\ .
\ee
From this we get the $\Gamma^0(4)$  monodromies in the $\Pi$ basis
\be
M_{\varphi=0}=\left(\begin{array}{rrr} 1& 0&0 \\ -1& -1& -4\\ 0&0&-1\end{array}\right),
\quad
M_{\varphi^2=-16}=\left(\begin{array}{rrr} 1& 0&0 \\ 0&3& 4\\
    0&-1&-1\end{array}\right),
\quad
M_{\varphi=\infty}=\left(\begin{array}{rrr} 1& 0&0 \\ 1&1& 0\\ -1&-1&1\end{array}\right) \ .
\ee
One checks $(M_{\varphi^2=-16} M_{\varphi=\infty})^{-1}=M_{\varphi=0}$.

In topological string theory or $N=2$ 4d supersymmetric
gauge theory, the coupling constants are complex.
At the various critical points one has to chose appropriate coordinates, which
are either invariant or reflect invariances of the
theory under the local monodromy. For example, at large
radius or the asymptotic free region of the gauge theory,
one can chose $T$ as the appropriate variable and the monodromy
$T \rightarrow  T+1$ is understood as a shift
in the NS $B$ field of topological string or the
$\theta$-angle of Yang-Mills theory. The canonical
choices of other coordinates in different regions
in the moduli space correspond to a change of
polarization.

Because in ABJM theory the coupling constant is real, there
is {\it a priori} no need to consider the action
of the monodromy. The polarization is picked once
and for all at the weak coupling point. The choice made
here is identical to the one made in topological string
theory at this point in moduli space. However, as pointed out in \cite{mpwilson}, 
this polarization is not the one of topological string theory at large radius. The coupling of
the ABJM theory $\lambda$ behaves at large radius
like
\be
\lambda=\partial_T F^0_{gw}-\frac{1}{2} T=
-\frac{1}{2}T^2-\frac{1}{2}T-\frac{1}{12}+{\cal O}(Q)\ .
\ee
To obtain the famous $N^{3/2}$
scaling of the genus zero free energy $F^{(0)}$, it is crucial to integrate
the $B$-cycle integral
$\partial_{\lambda} F^{(0)}$ with respect to $\lambda$~\cite{dmp}.
This yields\footnote{As further explained in~\cite{dmp}
it is natural to shift $\lambda$ and consider instead 
$ \hat \lambda =\lambda-\frac{1}{24}$.}
\be
\label{scaling}
F=g_s^{-2} F^{(0)}=
\frac{\pi \sqrt{2}}{3} k^2
{\hat \lambda}^{\frac{3}{2}} +
{\cal O}({\hat \lambda}^0,\re^{- 2 \pi \sqrt{2\hat \lambda}}) \ .
\ee
The relation of the topological string theory to the ABJM
theory at this point is therefore given by a change of
polarization.

What is remarkable is that, despite the fact that
the action of the monodromy does not have a clear interpretation in
ABJM theory, the higher genus contributions to the
partition function of the theory have the same
modular invariance under $\Gamma^0(4)$ that they have in topological string theory. One might speculate
that the monodromy at the strong coupling region
reflects an invariance
of the theory, so far not understood, which involves non-perturbative
effects. Note that this monodromy does not
change the leading $N^{3/2}$ behavior.
A related issue concerns the $1/6$ BPS Wilson loop vev
itself. This vev is obtained as an integral over the
${\cal C}$ cycle. However, the integral of the same
differential over the dual ${\cal B}$-cycle
has no interpretation in ABJM theory.
If the monodromy action had a meaning in ABJM theory, it
would mix the two types of cycles.

\subsection{Wilson loops in the geometric description}

The Wilson loop vevs have a genus expansion of the form,
\be
\langle W^{1/6, 1/2}_n\rangle=\sum_{g=0}^\infty g_s^{2g-1}\langle W^{1/6, 1/2}_n\rangle_g,
\ee
and of course the ABJM matrix model correlators (\ref{n-corr}) have the same type of expansion. The first term in this expansion 
corresponds to the genus zero or planar vev. The exact planar vevs of $1/2$ BPS and $1/6$ BPS Wilson loops (for winding number $n=1$) were obtained in
\cite{mpwilson}, from the exact solution of the ABJM matrix model at large
$N$. We will now review these results.

The planar limit of the matrix model is completely determined by the densities of eigenvalues in the cuts, which were also obtained explicitly in \cite{mpwilson}:
\be
\ba
\rho_1(X)\rd X &={1\over  2 \ri \pi^2 \lambda } \tan^{-1}\left[ {\sqrt{ \alpha X-1-X^2 \over \beta X +1 + X^2}} \right] {\rd X \over X},\\
\rho_2(Y)\rd Y &={1\over 2  \ri \pi^2 \lambda } \tan^{-1}\left[ {\sqrt{\beta Y +1 + Y^2\over  \alpha Y-1-Y^2 }} \right]{\rd Y \over Y},
\ea
\ee
where
\be
\alpha=a+{1\over a}, \quad \beta=b+{1\over b}.
\ee
These densities are normalized in such a way that their
integrals over the cuts are equal to one. It is a standard result in matrix model theory that planar correlators of the form (\ref{n-corr}) are given by moments of the eigenvalue densities,
\be
\label{vevp}
g_s^{-1}\langle \tr\, \re^{n\mu_i} \rangle_{g=0} = N \int_{\CC_1} \rho_1(X) X^n  \rd X.
\ee
Keeping track of the residue at $X=\infty$, analogously to
(\ref{residuaatinfinity}), we can write simpler expressions for
the densities which are valid in the compactified variables $x,y$.
The planar $1/6$ BPS Wilson loop vevs read in terms of those
\be
g_s^{-1}\langle W_n^{1/6}\rangle_{g=0}=\frac{k}{2 \pi^2} \int_{\CC_1} \mu_n,
\qquad
g_s^{-1}\langle \hat W_n^{1/6}\rangle_{g=0}=(-1)^n\frac{k}{2 \pi^2} \int_{\CC_2}
\mu_n\ .
\ee
The planar $1/2$ BPS Wilson loops is given by the $\gamma$-period, i.e. the residue at infinity,
\be
g_s^{-1} \langle W_n^{1/2}\rangle_{g=0}=\frac{k}{2 \pi^2} \oint_{\gamma} \mu_n\ .
\ee
Since the forms $\omega_n$ defined in (\ref{relationbetweendifferentials}) are not independent
elements of the cohomology of the curve,  one can relate all Wilson loop vevs to the integrals of $\mu_0$. Let us
denote by 
\be
R_n(\varphi)=\oint_\gamma \mu_n
\ee
the residue of $\mu_n$ at $x=\infty$. Then, we get a relation in
homology of the form 
\be
\label{differentialrelation}
{\cal L}_n\omega_0-\omega_n= \partial_\varphi R_n(\varphi) x^3 \rd x
\ee
where
\be
\quad {\cal L}_n=p_n^1(\varphi)\partial_\varphi+ p_n^0(\varphi)\ .
\ee
The coefficients $p_n^0(\varphi)$ and $p_n^1(\varphi)$ are polynomials in
$\varphi$ and can be obtained by the Griffiths reduction method. For the
first few we get,
\be
\begin{array}{rlrlrl}
p_1^0&=\frac{\varphi}{4},\ & p_1^1&=0,\ & R_1&=\frac{1}{2} \varphi,\ \\
p_2^0&=1,\ & p_2^1&=4 \varphi -\frac{\varphi^3}{4} , \ & R_2&=\frac{1}{4}
\varphi^2, \\
p_3^0&=\frac{9 \varphi}{4}-\frac{\varphi^3}{8}, \ &
p_3^1&=6\varphi^2-\frac{3\varphi^4 }{8},\ & R_3&=\frac{1}{2} \varphi-
\frac{\varphi^3}{6},\\
p_4^0&=1+\frac{10\varphi^2}{3}-\frac{5\varphi^4 }{24}, \  &
p_4^1&=\frac{16\varphi^2}{3}+7 \varphi^3 -\frac{11\varphi^5 }{24},\ &
R_4&=\varphi^2+\frac{\varphi^4}{8}\ . \\
\end{array}
\ee
This relates $\langle W_n^{1/6}\rangle_{g=0}$ to
$\lambda$, e.g.
\be
 \langle W_1^{1/6}\rangle_{g=0}=\frac{1}{4}\int \kappa \lambda(\kappa)
\rd\kappa +\frac{1}{2}\kappa\ .
\ee
The integration constant is zero as $\mu_1$ has no constant residue.

The relations (\ref{differentialrelation}) are homological relations. 
They imply a differential relation between the $\CB$-cycles integrals over
$\mu_n$ and $\partial_\lambda  F^0$. Since $\lambda$ and 
$\partial_\lambda F^0$ are related by special geometry, the relations
(\ref{differentialrelation}) imply, for each $n$, differential
relations between the Wilson loop integrals over the $\CC$ and the $\CB$ cycles. 
These can be viewed as an extension of special geometry to the Wilson
loop integrals.

We will now compute the vev (\ref{vevp}) for any positive integer $n$, at leading order in the strong coupling expansion, extending the result for $n=1$ 
obtained in \cite{mpwilson}. In the form (\ref{vevp}), these correlators are difficult to compute, but as in \cite{bt}, their derivatives w.r.t. $\kappa$ are easier to calculate and given by
\be
g_s^{-1}{\partial \over \partial \kappa} \langle W^{1/6}_n \rangle_{g=0} ={k \over 2 \pi^2 } \CI_n,
\ee
where
\be
\label{in-int}
\CI_n = \frac{1}{2}\int\limits_{1/a}^a\frac{X^n \rd X}{\sqrt{({\alpha X-1-X^2})({\beta X+1+X^2})}},
\ee
which can be calculated in terms of elliptic integrals. The computation for $n=1$ was done in \cite{mpwilson}, and in Appendix A we compute them for
a positive integer $n$. In order to make contact with the Fermi gas approach, where subleading exponential corrections are neglected, we want to extract their leading exponential behavior in the strong coupling region $\kappa \gg 1$. One finds,
\be
\label{leading-in}
\CI_n \approx {\ri^n \kappa^{n-1} \over 2}\left(  \log \kappa - {\pi \ri\over 2} - H_{n-1} \right), \qquad \kappa \gg 1,
\ee
where
\be
H_n =\sum_{d=1}^n {1\over d}
\ee
are harmonic numbers (for $n=1$, we set $H_0=0$). It then follows that
\be
\label{W16kappa}
g_s^{-1}\langle W^{1/6}_n \rangle_{g=0} ={( \ri  \kappa)^{n}  k \over 4 \pi^2 n } \left( \log \kappa - {\pi \ri \over 2} - H_{n} \right)\left(1 +{\CO}\left({1\over \kappa^2} \right) \right).
\ee
From this we deduce that, for the $1/2$ BPS Wilson loop, one has
\be
g_s^{-1}\langle W^{1/2}_n \rangle_{g=0} = -{\ri k (\ri \kappa)^n  \over 4  \pi n}  \left(1 +{\CO}\left({1\over \kappa^2} \right) \right).
\ee
This agrees with a result obtained in section 8.2 of \cite{dmp}, where the generating function of these vevs, with an extra $1/n$ factor, was shown to be a dilogarithm in the variable $\ri \kappa$.

The regime of large $\kappa$ corresponds to the regime of large 't Hooft coupling \cite{mpwilson}, and one has from (\ref{lamkap}),
\be
\label{kappalambda}
 \lambda(\kappa) ={\log ^2(\kappa)\over 2 \pi
   ^2}+\frac{1}{24}+\CO\left( {1\over \kappa^2}\right),
   \ee
   which is immediately inverted to
   \be
   \kappa=\re^{\pi {\sqrt{2\lambda}}}\left(1 +\CO\left(\re^{-2\pi {\sqrt{2\lambda}}} \right) \right).
   \ee
It follows that the $1/6$ Wilson loops go like,
\be
\langle W^{1/6}_n \rangle \approx \re^{n\pi {\sqrt{2\lambda}}},
\ee
and for $n=1$ this is in agreement with the AdS calculation in terms of fundamental strings \cite{dp,cw,rey}.

\subsection{The higher genus calculation from the spectral curve}
\label{thooftwilson}

The essential information of the higher genus expansion is
encoded in the expansion of the resolvent
\be
\mu_0(x)= \sum_{g=0}^\infty g_s^{2g} \mu^{(g)}_0(x)\ .
\label{genusexpansionmu0}
\ee
The densities for the Wilson line  integrals of winding $k$
at genus $g$ are then
\be
\mu^{(g)}_k(x)=x^k \mu^{(g)}_0(x)\ .
\label{higherwindinghighergenus}
\ee
The main task is hence to determine the
expansion (\ref{genusexpansionmu0}). To do this, we will use the matrix model
recursion of \cite{eynard,eo}, and we will present results and formulae which are valid for any spectral curve of genus one. We will then specialize to the 
spectral curve describing ABJM theory. 

The simplest formulation of the topological recursion uses the
hyperelliptic curves
\be
y^2=\sigma(x)
\label{hyperelliptic2}
\ee
and as meromorphic differential defining the filling fractions
\be
\Phi(p)=y(p) \rd x (p)\ .
\label{standardfillingfraction}
\ee
Instead of (\ref{standardfillingfraction}) we want
to work with $\mu_0$ as differential defining the
filling fractions, as in \cite{mmopen,bkmp}. Let us denote the points on the
$\pm$-branch  of (\ref{hyperelliptic}) $p$ and $\bar p$, i.e. both points map to the same $x$ value. In the
recursive formalism of \cite{eo} the discontinuity of $\Phi(p)-\Phi(\bar p)$
over the cuts is essential. Likewise at $\sigma(X)=0$ the two
branches of the curve (\ref{Y}) come together.
The difference of $\mu_0$ on the two branches is
however
\be
\mu^{(0)}_k(p)-\mu^{(0)}_k(\bar p)= \frac{2}{X^{k-1}} {\rm tanh}^{-1}
\left(\frac{\sqrt{\sigma(X)}}{a(X)}\right){\rm d} X \ .
\label{differenceonsheets}
\ee
One can now redefine $y$ in order to match these differences.
This leads to the definition of a curve
\be
{\tilde y}^2=M^2(x)\sigma(x)\ ,
\ee
on which  (\ref{standardfillingfraction}) is equivalent
to $\mu_0$ on (\ref{mirrorcurvep1xp1b},\ref{Y}).
Note that the resulting moment function \cite{mmopen,bkmp}
\be
M(x)=\frac{2}{x \sqrt{\sigma(x)}} {\rm tanh}^{-1}
\frac{\sqrt{\sigma(x)}}{a(x)}\
\ee
does not modify the branch points. In particular it does
not introduce new ones. 

The recursion formula of \cite{eynard, eo} reads\footnote{One writes
$W_g(p_1,\ldots,p_k)\rd p_1\ldots \rd p_k:=\omega_g(p_1,\ldots,p_k)=\mu_0^{(g)}(p_1,\ldots,p_k)$.}
\be
W_g(p,p_1,\ldots, p_k)=\sum_{i} {\rm Res}_{q=x_i} \frac{\rd S(p,q)}{\tilde y(q)}
\left( \sum_{h=0}^g \sum_{J\subset K} W_h(q,p_J) W_{g-h}(q,p_{K\setminus J} )+
  W_{g-1}(q,q,p_K)\right)\ .
\label{recursion}
\ee
Here $K,J$ are  index sets  $K=\{ 1,\ldots, k\}$ etc. In principle
we are only interested in the $W_g(p)$, however for $g=2$ the recursion requires to calculate amplitudes with up to
three legs at genus 0. The $\rd S(p,q)$ are the unique meromorphic
differentials with only simple poles at $q=p$ and  $q=\bar p$, whose
integral w.r.t. to $q$  over the $A$ cycles, which we call in our context
${\CC}$,  vanish,
\be
\rd S(q,p) \ _{ \sim \atop ^{q\rightarrow p}}\  \frac{\rd q}{q-p},\qquad \rd S(q,p) \
_{ \sim \atop ^{q\rightarrow \bar p}}\  - \frac{\rd q}{q-p},\qquad \int_{q\subset {\cal
    A}_i} \rd S(q,p)=0  \ .
\label{conditionsondS}
\ee
The decisive technical tools to solve the recursion are
the so called kernel  differentials (see for example \cite{akemann,eynard})
\be
\chi_i^{(n)}= {\rm Res}_{q=x_i}\left(\frac{\rd S(p,q)}{\tilde y(q)}
\frac{1}{(q-x_i)^n}\right)\ .
\label{defkerneldiff}
\ee
Multiplying an expression $f(q,p_i, x_i)$ by  $\frac{\rd S(p,q)}{y(q)}$
and taking the sum of the residua  at $q=x_i$ is the crucial step
in solving the recursion, so let us denote
\be
\Theta(p,q) f(q,p_i, x_i)=\sum_{i} {\rm Res}_{q=x_i} \left(\frac{\rd
    S(p,q)}{\tilde y(q)} f(q,p_i, x_i)\right) \ .
\ee

Besides the genus zero resolvent $\mu_0(x)$, in order to start the recursion
one needs the annulus amplitude,
\be
W_0(p,q)= -\frac{1}{2(p-q)^2}+ \frac{\sigma(p)}{2(p-q)^2
  \sqrt{\sigma(p)} \sqrt{\sigma(q)}} - \frac{\sigma'(p)}{4(p-q)
  \sqrt{\sigma(p)} \sqrt{\sigma(q)}} + \frac{A(p,q)}{\sqrt{\sigma(p)}
  \sqrt{\sigma(q)}}\ .
\label{annulus}
\ee
which is related to the Bergman kernel by
\be
B(p,q)=\left(W_0(p,q)+\frac{1}{(p-q)^2}\right) \rd p \rd q \ .
\ee

On an elliptic curve $A(p,q)$ as well as the kernel
differentials  are given in terms of elliptic
integrals:
\be
A(p,q)= (p-x_1) (p-x_2) +(p-x_3) (p-x_4) +(x_1-x_2) (x_4-x_2)G(k)
\ee
where
\be
k^2=\frac{(x_1-x_2)(x_3-x_4)}{(x_1-x_3)(x_2-x_4)}
\ee
is the elliptic modulus and
\be
\label{gk}
G(k)=\frac{E(k)}{K(k)}
\ee
is the ratio between the two complete elliptic integrals
\be E(k)=\int_{0}^{\frac{\pi}{2}} \sqrt{1- k^2 \sin^2 \theta }\rd \theta,\qquad
K(k)=\int_{0}^{\frac{\pi}{2}} \frac{ \rd \theta}{ \sqrt{1- k^2 \sin^2 \theta}}
\ . \
\ee

As explained in~\cite{kmt} the ordering of the branch points here
follows the one appropriate for local $\mathbb{P}^1\times \mathbb{P}^1$,
which is obtained from the one in \cite{akemann} by the exchange
\be
x_2 \leftrightarrow x_4 \ .
\ee
The expression  for the kernel differentials follows from a Taylor expansion
of
\be
\frac{\rd S(p,q)}{\tilde y(q)}=
\frac{1}{M(q)\sqrt{\sigma(p)}}\left(\frac{1}{p-q}+{\cal N}^{(1)}(q)\right)d p\
\ee
around the branch points.
Here
\be
{\cal N}^{(1)}(q)={\cal K} C^{(1)}(q)=\frac{\pi \sqrt{(x_1-x_3)(x_2-x_4)}}{2 K(k)}
C^{(1)}(q)
\label{normalisation}
\ee
is a normalization of the ${\cal C}$ (or equivalently the  ${\cal A}$) cycle
integral
\be
C^{(1)}(q)=\int_{\cal C}\frac{1}{2 \pi \ri} \frac{dx}{(q-x) \sqrt{\sigma}}\ ,
\ee
so that the last property (\ref{conditionsondS}) hold.
Note that, if $q$ approaches the
branch points of the cuts defining the ${\cal C}$ cycle,  this integral has to be regularized, 
\be
C^{(1)}(x_i)=\begin{cases}
\left.\frac{1}{2\pi \ri}\int_{\cal C} \frac{\rd x}{(q-x) \sqrt{\sigma}}\right|_{q=x_i}& \text{if } x_i
\text{ is not a branch point defining } {\cal C}  \\
\left.\frac{1}{2\pi \ri} \int_{\cal C} \frac{\rd x}{(q-x) \sqrt{\sigma}}
  -\frac{1}{\sqrt{\sigma(q)}}\right|_{q=x_i}
& \text{if } x_i \text{ is a branch point defining } {\cal C}
\end{cases}
\ee

This definition of the regularization ensures that one
can move the contour from the $x_1-x_2$ cut to the $x_3-x_4$ cut
without getting a contribution from the poles. As a
consequence the so defined integrals $C(x_i)$ obey
a symmetry under certain permutations of the branch
points. We can evaluate e.g. the manifestly regular integral\footnote{Here we
  make contact with the shorthand  notation $\alpha_i$ introduced
  in~\cite{akemann}.}
\be
\alpha_4={\cal N}^{(1)}(x_4)=\frac{1}{x_4-x_3}
\left[\frac{(x_3-x_1)}{(x_1-x_4)} G(k)+ 1\right]
\ee
and obtain from the symmetrization the evaluation at the other
branch points
\be
{\cal N}^{(1)}(x_1)={\cal N}^{(1)}(x_4)|_{x_1\leftrightarrow x_4\atop x_2\leftrightarrow
  x_3}, \quad {\cal N}^{(1)}(x_2)={\cal N}^{(1)}(x_4)|_{x_1\leftrightarrow
x_3\atop x_2\leftrightarrow x_4}, \quad {\cal N}^{(1)}(x_3)={\cal N}^{(1)}(x_4)|_{x_1\leftrightarrow x_2\atop
x_3\leftrightarrow x_4} \ .
\label{symmetrizations}
\ee

Higher kernel differentials are therefore given by
\be
\chi_i^{(n)}=\frac{1}{(n-1)!} \frac{1}{\sqrt{\sigma(q)}}
\frac{\rd^{n-1}}{\rd q^{n-1}} \left[\frac{1}{M(q)} \left(\frac{1}{p-q} +{\cal N}^{(l)}(q)\right)\right]_{q=x_i} \ .
\ee
Here  ${\cal N}^{(l)}(q)={\cal K}C^{(l)}(q)$, and since the normalization
factor ${\cal K}$ is independent of $q$, the only nontrivial task is to
calculate the derivatives 
\be
C^{(n)}(q)=\frac{\rd^{n-1}}{\rd q^{n-1}} C^{(1)}(q).
\ee
There are various ways to do this. One fast way is to compute
\be
C^{(n)}(q)=\frac{(-1)^{n-1} (n-1)!}{2 \pi i}\int_{\cal C} \frac{\rd x}{(q-x)^n
  \sqrt{\sigma}} \ .
\ee
These integrals have poles at finite points and are very
similar to the ones with poles at infinity. By similar
formulas they can be expressed by linear expressions in
$K(k)$ and $E(K)$ with rational coefficients in the moduli.
In particular the normalized integrals ${\cal  N}^{(n)}(q)$
depend only on the ratio of elliptic functions $G(k)$ defined in (\ref{gk}). 
To get expressions which are valid at all branch points
one calculates first ${\cal N}^{(n)}(x_4)$, which is regular, and then uses
(\ref{symmetrizations}) to get ${\cal N}^{(n)}(x_i)$.
These derivatives have symmetric expressions in terms of
the branch points and the $\alpha_i$. E.g. the first two derivatives
are
\be
\begin{array}{rl}
{\cal N}^{(2)}(x_i)&=\displaystyle{\frac{1}{3}\sum_{j\neq i}
\frac{\alpha_j-\alpha_i}{x_j-x_i}}\ , \\ [ 3 mm ]
{\cal N}^{(3)}(x_i) &=\displaystyle{\frac{2}{15}\left[\frac{1}{\prod_{j\neq i} (x_j-x_i)}+  \sum_{j\neq i}\left(
\frac{7 \alpha_j-\alpha_i}{(x_j-x_i)^2}+3\sum_{j\neq k}
\frac{1}{(x_j-x_i)^2(x_k-x_i)}\right)\right]}\ . \\
\end{array}
\ee

Eventually one needs integrals over meromorphic forms with mixed poles
\be
\omega_{n,k}= \frac{x^n}{(x-p)^k \sqrt{\sigma(x)}} \rd x\ ,
\ee
which are obtained from the obvious relations
\be
\omega_{n,k}=\omega_{n,k-1}+p\omega_{n-1,k}\ .
\label{trivial}
\ee

The genus one differential is then determined by evaluating
\be
W_1(p)= \Theta(p,q) W_0(q,q)\
\ee
using (\ref{annulus},\ref{defkerneldiff}), as well as the explicit
formulas for the kernel differentials for elliptic curves. It was first calculated explicitly in \cite{akemann}. One can order $W_1(p)$ according to its poles at the
branch points
\be
W_1(p)=\frac{4}{\sqrt{\sigma(p)}} \sum_{i=1}^{4}\left(\frac{A_i}{(p-x_i)^2}+
 \frac{B_i}{p-x_i} + C_i\right)\ ,
\label{laurantW1}
\ee
where
\be
\ba
A_i&=\frac{1}{16}\frac{1}{M(x_i)},\qquad
B_i=-\frac{1}{16}\frac{M'(x_i)}{M^2(x_i)}+\frac{1}{8 M(x_i)}\left( 2 \alpha_i-\sum_{j\neq i} \frac{1}{x_i-x_j}\right)  \ ,\\
C_i&= -\frac{1}{48}\frac{1}{M(x_i)}\sum_{j\neq i}\frac{\alpha_i-\alpha_j}{x_j-x_i}  -\frac{1}{16}\frac{M'(x_i)}{M^2(x_i)}\alpha_i+\frac{\alpha_i}{8 M(x_i)}\left( 2
  \alpha_i-\sum_{j\neq i} \frac{1}{x_i-x_j}\right) \ .\\
\ea
\label{laurant}
\ee
To obtain the form $\omega_2(p)$
\be
W_2(p)= \Theta (p,q) \left( W_1(q,q) + W_1(q) W_1(q)\right) \
\ee
one needs $W_1(p,p_1)$ from
\be
W_1(p,p_1)=  \Theta (p,q) \left( W_0(q,q,p_1)+
2 W_1(q) W_0(q,p_1)\right) \
\ee
and $W_0(p,p_1,p_2)$ from
\be
W_1(p,p_1,p_2)= 2\Theta (p,q) W_0(q,p_1)
W_0(q,p_2) \ .
\ee
By repeated application of the recursion, one
expresses any amplitude through a calculation 
of repeated residues of products of the annulus
amplitude. E.g.
\be
\ba
W_2(p)=& 2 \Theta( p,q) \Theta(q,q_1) \Theta(q_1,q_2) W_0(q_2, q) W_0(q_2, q_1)+ \\
       &2 \Theta(p,q) \Theta(q,q_1) \Theta(q_1,q_2) W_0(q_1, q) W_0(q_2, q_2)+\\
      & \Theta(p,q) \Theta(q,q_1) \Theta(q,q_2) W_0(q_1, q_1) W_0(q_2, q_2)\ .
\ea
\ee
It is easy to derive that for amplitude with  genus $g$ and $h$
holes all terms will be of the general
form
\be
W_{g,h}\sim \Theta^{2g-2+h} W^{g+h-1}_{0,2}\ .
\ee
However the number of terms grow exponentially with $g$ and $h$.
A few examples for the number of contributions counted
with multiplicity is given in the table below.
\begin{table}[h!]
\centering
\begin{tabular}[h]{|c|cccccccc|}
\hline
      & $g$ & 0  &  1     &  2 &  3 &   4 &    5 & \\
\hline
$h$   &   &         &    &    &       &      &    &        \\
1     &  & disk       &     1   &  5 & 60 &  1105 & 27120  &   \\
2     &  & 1          &     4    & 50  &  960 &   24310 &  &  \\
3     &  & 2           &    32   & 700  & 19200&   & &  \\
4     &  & 12          &   384   & 12600&      &   & &  \\
\hline
\end{tabular}
\caption{Number of terms involved in the recursive definition of
$W_{g,h}$.}
\end{table}

Since $W\sim G$, and each $\Theta$ increases
the power of $G$ by one, we get for the leading power $W_{g,h}\sim G^{3g+ 2h-3}$.
More precisely, the $W_{g,h}$ are meromorphic differentials with the
following pole structure
\be W_{g,h}(p_1,\ldots, p_h) =\frac{1}{\prod_l^h \sqrt{\sigma(p_l)} }
\left(\sum_{j=0}^{3g-2+h}\sum_{k=1}^h \sum_{i=1}^4 \frac{A^{(j)}_{i,k}}{(p_k-x_i)^{3g-2+h-j}}\right)\ ,
\ee
where
\be
A^{(j)}_{i,k}=\sum_{p=0}^j G^p a^{(p)}_{i,k}(x_i)
\ee
are polynomials in the ratio of the complete elliptic integrals. 
For $W_g(p)$, $g=2,3$ we found a explicit expressions for 
general moment functions. To write down all  $A^{(j)}_{i,k}$ takes however
several pages. We display the coefficient of the leading pole
\be
A^{(0)}_{i,1}=\frac{105}{2^7 M(x_i)^3 \prod_{k\neq i} (x_i-x_k)}\ .
\ee
and $a^{(3g-2+h)}_1:=\sum_{i} a^{(3g-2+h)}_{i,1}$  multiplying the
highest power of $G^{3g-2+h}$ in $W_{g,h}(\underline{p})$. 
For $h=1$ we find   
\be
a^{(3g-2+1)}_1=c_{g,1}\left(\sum_{i=1}^4\frac{1}{M(x_i) \prod_{j\neq i} (x_i-x_j)^2}\right)^{2g-1}[(x_1-x_3)(x_2-x_4)]^{3g-1},     
\ee
where $c_{2,1}=-\frac{5}{16}$, $c_{3,1}=\frac{7}{8}$. The other expressions are available on request. 

All the results above are valid for any spectral curve of genus one. Let us particularize 
them for ABJM theory. The calculation of the higher genus functions $W_g(p)$ is obviously quite involved.
Nevertheless one can make a general statement about the
logarithmic structure of Wilson loop integrals at
strong coupling. Since
\be
G(\kappa)\sim \frac{1}{\log(\kappa)}+{\cal  O}(\kappa^{-1})
\ee
we see that the highest inverse powers
of $1/\log(\kappa)$  at leading order in $\kappa$
go as
\be
{1\over (\log(\kappa))^{3g-1}}.
\ee
For the $1/6$
Wilson loop there will be a positive power of
$\log(\kappa)$ at leading order due to the integration
of the meromorphic differential $\omega_g(p)$ over the ${\cal C}$
cycle. The structure can be checked e.g. at genus two, in the expression obtained from the Fermi gas approach in (\ref{genus2kappa}).
Using now (\ref{laurantW1}), one obtains the
weak coupling expansion of the 1/2 BPS Wilson line at genus one,
\be
\langle W^{1/2}_{n=1}\rangle_{g=1}= -\frac{1}{6} \ri \pi \lambda-\frac{11}{36} \ri \pi
^3 \lambda^3+\frac{97}{360} \ri \pi ^5 \lambda^5-\frac{10331 \ri \pi ^7
  \lambda^7}{30240}+{\cal O}\left(\lambda^{9}\right), 
\ee
which was already calculated in \cite{dmp} with the same procedure. The results presented above allow us to 
find the weak coupling expansion also at genus $2$, 
\be
\langle W^{1/2}_{n=1}\rangle_{g=2}=
\frac{7 {\rm i} \pi  \lambda }{5760}+\frac{29 \ri \pi ^3 \lambda^3}{2160}-\frac{7073 \ri \pi ^5 \lambda^5}{691200}-\frac{20077 \ri \pi ^7 \lambda^7}{1036800}+\frac{109387361 \ri \pi ^9 \lambda^9}{1045094400}+O\left(\lambda^{11}\right)\ .
\ee
The first few terms in this expansion have been checked against perturbative calculations in the matrix model. 

\sectiono{The Fermi gas approach}

\subsection{Quantum Statistical Mechanics in phase space}

The Fermi gas approach to the ABJM matrix models (and to other matrix models arising in $\CN=3$ Chern--Simons--matter theories) is
based on an exact equivalence with a quantum Fermi gas of $N$ particles with Planck constant $\hbar=2 \pi k$, and an evaluation of the
different observables in the semiclassical expansion. For this reason, it is convenient to formulate the quantum mechanical problem
in Wigner's formalism. We will now review some of the basic tools that we need to set up the formalism.

We recall that, to construct the Hilbert space for a space of indistinguishable particles, one introduces a projection operator on totally (anti)symmetric states
\be
P_\eta={1\over N!} \sum_{\sigma \in S_N} \eta^{\epsilon(\sigma)} \sigma,
\ee
where
\be
\eta=\pm 1
\ee
for bosons and fermions, respectively. This operator satisfies
\be
P_\eta^2=P_\eta.
\ee
Let
\be
| \lambda_1, \cdots, \lambda_N \rangle
\ee
be the basis of space eigenstates for an $N$-particle system $\CH_N$ of distinguishable particles. The appropriately (anti)symmetrized states
\be
\left| \lambda_1, \cdots, \lambda_N \right \} ={\sqrt{N!}} P_\eta | \lambda_1, \cdots, \lambda_N \rangle={1\over {\sqrt{N!}}}  \sum_{\sigma \in S_N} \eta^{\epsilon(\sigma)}
| \lambda_{\sigma(1)}, \cdots, \lambda_{\sigma(N)} \rangle
\ee
constitute are a basis of the Hilbert space of bosons/fermions $\CB_N$, $\CF_N$. The resolution of the identity in $\CB_N/\CF_N$ reads
\be
\label{resid}
{1\over N!} \int \rd \lambda  \left| \lambda_1, \cdots,  \lambda_N\right\}\left\{ \lambda_1, \cdots, \lambda_N \right| ={\bf 1}.
\ee

A $n$-body operator $\CO$ is an operator which is invariant under any permutation of the particles, and acts on a state of $\CH_N$ as follows,
\be
\CO |\lambda_1 \cdots \lambda_N \rangle = {1\over k!} \sum_{1\le i_1 \not= \cdots \not= i_k \le N} \CO(\lambda_{i_1}, \cdots, \lambda_{i_k})|\lambda_1 \cdots \lambda_N \rangle
\ee
For example, for a one-body operator we simply have
\be
\CO |\lambda_1 \cdots \lambda_N \rangle = \sum_{i=1}^N  \CO(\lambda_i)|\lambda_1 \cdots \lambda_N \rangle,
\ee
where $\CO(\lambda)$ is an operator on the Hilbert space of a single particle.

In the canonical ensemble, the thermodynamic properties of the system are encoded in the canonical density matrix. For a system of distinguishable particles,
the canonical density matrix is given by
\be
\label{drho}
\rho_D (\{ x_1, \cdots, x_N\} , \{x_1',\cdots, x'_N\} ;\beta)=\langle x_1 \cdots x_N | \re^{-\beta \hat H} | x'_1 \cdots x'_N \rangle,
\ee
where $\hat H$ is the total Hamiltonian of the $N$ particles. For bosons (respectively, fermions), we have to (anti)symmetrize it in an appropriate way, to obtain \cite{feynman}
\be
\ba
\rho (\{ x_1, \cdots, x_N\} , \{x_1',\cdots, x'_N\} ;\beta)&={1\over N!} \sum_{\sigma \in S_N} \eta^{\epsilon(\sigma)} \rho_D (\{ x_1, \cdots, x_N\} , \{x_{\sigma(1)}',\cdots, x'_{\sigma(N)}\} ;\beta)\\
&={1\over N!} \{ x_1 \cdots x_N | \re^{-\beta \hat H} | x'_1 \cdots x'_N \}.
\ea
\ee

In order to compute the vevs of many-body operators in the canonical ensemble,
it is useful to introduce density submatrices or reduced density matrices (see for example \cite{feynman,zuk}).
The reduced $n$-particle density matrix is defined as
\be
\label{reduze}
\rho_n (\{ x_1, \cdots, x_n\} , \{x_1',\cdots, x'_n\} ;\beta)={N!\over (N-n)!}  \int \rd x_{n+1}\cdots \rd x_N \, \rho(\{ x_1, \cdots, x_N\} , \{x_1',\cdots, x'_N\} ;\beta)
\ee
The thermal average of an $n$-body operator $\CO$ in the canonical ensemble is defined by
\be
\label{unvev}
\langle \CO \rangle_N = \tr\left( \hat \rho \CO\right)
\ee
where we are using {\it unnormalized} vevs. This can be computed in terms of the $n$-reduced density matrix
as follows
\be
\label{unvevdens}
\langle \CO \rangle_N={1\over n!} \int \rd x_1 \cdots \rd x_n \CO (x_1, \cdots, x_n) \rho_k(\{ x_1, \cdots, x_n\} , \{x_1',\cdots, x'_n\} ;\beta).
\ee
In our conventions, the canonical partition function is defined as the thermal average of the identiy,
\be
Z_N=\tr(\rho).
\ee
We note that, in a system of non-interacting particles, the density matrix (\ref{drho}) factorizes,
\be
\rho_D (\{ x_1, \cdots, x_N\} , \{x_1',\cdots, x'_N\} ;\beta) =\prod_{i=1}^N \rho(x_i, x'_i),
\ee
where $\rho(x,x')$ is the canonical density matrix of the one-particle problem.

In many situations it is more useful to work in the grand-canonical ensemble, where the reduced density matrix is defined as (see for example \cite{zuk})
\be
\rho_n^{\rm GC} (\{ x_1, \cdots, x_n\} , \{x_1',\cdots, x'_n\} ;\beta, z)= \sum_{N=n}^{\infty} z^N \rho_n (\{ x_1, \cdots, x_n\} , \{x_1',\cdots, x'_n\} ;\beta)
\ee
and as usual
\be
z=\re^{\beta \mu}
\ee
denotes the fugacity. The grand partition function is
\be
\Xi= 1+\sum_{N=1} z^N Z_N,
\ee
and the vev of an $n$-body operator in this ensemble can be simply expressed in terms of a sum of canonical
vevs over all particle numbers,
\be
\langle \CO \rangle^{\rm GC}=\sum_{N=n}^{\infty} \langle \CO \rangle_N z^N.
\ee

In the case of non-interacting gases, the grand-canonical density matrix has a very simple form (see for example \cite{zuk,hara}):
\be
\label{kGC}
\rho^{\rm GC}_n (\{ x_1, \cdots, x_n\} , \{x_1',\cdots, x'_n\} ; \beta, z)=\Xi \, \sum_{\sigma \in S_n } \eta^{\epsilon(\sigma)} \prod_{i=1}^n n(x_i, x'_{\sigma(i)}; \beta, z),
\ee
where $\Xi$ is the grand-canonical partition function, and
\be
n(x, x'; \beta, z)=\left\langle x\left| {1\over z^{-1} \re^{\beta \hat H}-\eta} \right| x'\right\rangle
\ee
is the occupation number operator in the position representation. The relationship (\ref{kGC}) can be derived by using creation and annihilation operators \cite{hara}.
There is also an elegant derivation in the case $n=1$ by using the so-called Landsberg's recursion relation. This relation is based on the analysis of the sum over permutations in terms of
conjugacy classes, and it  was originally derived for the canonical partition function of ideal
quantum gases (see for example \cite{krauth}). It is however straightforward to generalize it to density matrices \cite{ck}, and one finds
\be
\rho_1(x,x';\beta)= \sum_{\ell=1}^N\eta^{\ell-1} \rho(x,x'; \ell \beta) Z_{N-\ell},
\ee
where $\rho(x,x';\beta)$ is the density matrix for the one-particle problem. We now sum over all $N$ with the fugacity $z^N$ to obtain the grand-canonical,
reduced density matrix,
\be
\ba
\rho_1^{\rm GC}(x,x';\beta,z)&=\sum_{N=1}^{\infty} \rho_1(x,x';\beta) z^N=\eta \sum_{N=1}^{\infty} \sum_{\ell=1}^N Z_{N-\ell} z^{N-\ell}\left \langle x\left| \re^{-\ell \beta \hat H} \right |x'\right\rangle (\eta z)^\ell \\
&=\eta \left( \sum_{M=0}^{\infty}Z_M z^M \right)\left \langle x\left | \sum_{\ell=1}^{\infty} \left( \eta z \right)^\ell \re^{-\ell \beta \hat H }\right |x'\right\rangle \\
&=\Xi \,  \left \langle x\left| {1\over z^{-1} \re^{\beta \hat H}-\eta} \right|x'\right\rangle
\ea
\ee
We conclude in particular that the vev of a one-body operator in the grand-canonical ensemble is given by
\be
\label{GCvev}
\langle \CO \rangle^{\rm GC}=\Xi \tr \left(  {\CO \over z^{-1} \re^{\beta \hat H}-\eta} \right)
\ee
where the operator $\CO$ appearing inside the trace is understood as the operator restricted to the one-particle Hilbert space.

In order to calculate the quantum-mechanical averages, we will use a semiclassical or WKB expansion. The most convenient framework to do this
is the phase space formulation of Quantum Mechanics (see \cite{hillery,cz} for detailed expositions).
We first recall that the Wigner transform of an operator $\hat A$ is given by%
\be
\label{wignert}
A_{\rm W}(q,p)=\int \rd q' \left\langle q-{q'\over 2}\right|\hat A \left| q+{q'\over 2}\right\rangle \re^{\ri p q'/\hbar}.
\ee
The Wigner transform of a product is given by the $\star$-product of their Wigner transforms,
\be
\label{starprod}
\left(\hat A \hat B\right)_{\rm W}=A_{\rm W}\star B_W
\ee
where the star operator is given as usual by
\be
 \star=\exp\left[ {\ri \hbar \over 2} \left( {\overleftarrow{\partial}}_q {\overrightarrow{\partial}}_p  - {\overleftarrow{\partial}}_p {\overrightarrow{\partial}}_q\right) \right],
 \ee
and is invariant under linear canonical transformations. Another useful property is that
 \begin{equation}
 \label{trA}
  \Tr \hat{A}=\int\frac{\rd p \rd q }{2\pi\hbar}\,A_\wigner(q,p).
 \end{equation}

 Let $\hat H$ be the Hamiltonian of a one-particle, one-dimensional
 quantum system, and let $H_{\rm W}$ be its Wigner transform. Following \cite{gramvoros} we notice that it is possible
 to expand any function $f(\hat H)$ of $\hat H$ around $H_{\rm W}(q,p)$, which is a $c$-number. This gives,
\be
f(\hat H) = \sum_{r\ge 0} {1\over r!} f^{(r)}( H_{\rm W}) \left( \hat H -H_{\rm W}(q,p)\right)^r.
\ee
The semiclassical expansion of this object is obtained simply by evaluating its Wigner transform, and we obtain
\be
\label{wfH}
f(\hat H)_{\rm W} = \sum_{r\ge 0} {1\over r!} f^{(r)}\left( H_{\rm W} \right)\CG_r
\ee
where
\be
\label{Gr}
\CG_r=\left[ \left( \hat H -H_{\rm W}(q,p)\right)^r \right]_{\rm W}
\ee
and the Wigner transform is evaluated at the same point $q,p$. Of course, one has
\be
\CG_0=1, \qquad \CG_1=0,
\ee
and the quantities $\CG_r$ for $r\ge 2$ can be computed by using (\ref{starprod}). They have an $\hbar$ expansion of the form
\be
\label{GrEx}
\CG_r=\sum_{n\ge \left[\frac{r+2}{3}\right]} \hbar^{2n} \CG_r^{(n)},  \qquad r\ge 2.
\ee
This means, in particular, that to any order in $\hbar^2$, only a finite number of $\CG_r$'s are involved.
One finds, for the very first orders \cite{gramvoros},
\be
\ba
\CG_2&=-{\hbar^2 \over 4} \left[ {\partial^2 H_{\rm W} \over \partial q^2}{\partial^2 H_{\rm W} \over \partial p^2}-\left( {\partial^2 H_{\rm W} \over \partial q \partial p}\right)^2 \right] +\CO(\hbar^4), \\
 \CG_3&=-{\hbar^2 \over 4} \left[\left( {\partial H_{\rm W} \over \partial q} \right)^2  {\partial^2 H_{\rm W} \over \partial p^2}+
 \left( {\partial H_{\rm W} \over \partial p} \right)^2  {\partial^2 H_{\rm W} \over \partial q^2}-2  {\partial H_{\rm W} \over \partial q} {\partial H_{\rm W} \over \partial p}{\partial^2 H_{\rm W} \over \partial q \partial p}  \right]+\CO(\hbar^4).
 \ea
 \ee
One can then apply this method to compute the semiclassical expansion of any function of the Hamiltonian operator. A particularly important
operator is the distribution operator
\be
\label{Nop}
\hat n(E)=\Theta(E -\hat H)\ ,
\ee
where $\Theta(x)$ is the Heaviside step function. The trace of this operator gives the function $n(E)$, counting the number of
eigenstates whose energy is less than $E$:
\be
\label{Noper}
n(E)=\tr \, \hat n(E) =\sum_n  \Theta(E -E_n).
\ee
One can regard also the operator (\ref{Nop}) as the Fermi occupation number operator in the limit of zero temperature. When we apply (\ref{wfH}) to (\ref{Nop}),
we find,
\be
\hat n(E)_{\rm W}= \Theta(E-H_{\rm W})+ \sum_{r=2}^{\infty} {(-1)^r \over r!} \CG_r \delta^{(r-1)}(E-H_{\rm W}),
\ee
and evaluating the trace according to (\ref{trA}) one obtains the useful formula,
\be
\label{exactdens}
n(E)=\int_{H_{\rm W}(q,p) \le E} {\rd q \rd p \over 2 \pi \hbar} +  \sum_{r=2}^{\infty} {(-1)^r\over r!} \int {\rd q \rd p \over 2 \pi \hbar} \CG_r \delta^{(r-1)}(E-H_{\rm W}).
\ee
When (\ref{wfH}) is applied to the canonical density matrix at inverse temperature $\beta$, one finds,
\be
\label{genwk}
\left( \re^{-\beta \hat H} \right)_{\rm W}=\left( \sum_{r=0}^{\infty}{ (-\beta)^r  \over r!} \CG_r \right) \re^{-\beta H_{\rm W}}.
\ee
We will call the the functions $\CG_r$ appearing in (\ref{wfH}) {\it Wigner--Kirkwood corrections}, and the resulting expansions {\it Wigner--Kirkwood expansions}. These corrections were originally introduced by Wigner and Kirkwood in their study of the semiclassical expansion (\ref{genwk}) of the canonical partition function. Note that (\ref{genwk}) can be interpreted as saying that the Wigner transform of the canonical density matrix is the generating function of the Wigner--Kirkwood corrections. This will be useful later on.

Let us now come back to the calculation of statistical-mechanical averages. We will organize their calculation in two steps: first we will perform a low-temperature expansion (which is nothing but the Sommerfeld expansion used in the theory of free Fermi gases), expressing the finite temperature average in terms of a zero-temperature average. Then, we will evaluate this zero-temperature
average by using the Wigner--Kirkwood expansion. The reason we use the low-temperature expansion is that, in the thermodynamic system relevant to ABJM theory, large $N$ means large $\mu$ and large $E$, and this is equivalent to large $\beta$, i.e. low temperature.

We will focus on the one-body average appearing in (\ref{GCvev}), and we will restrict now to Fermi systems (i.e. we set $\eta=-1$).
We first recall that the Sommerfeld expansion expresses any integral of the form
\be
I=\int_0^\infty {g(E) \over \re^{\beta(E-\mu)} +1} \rd E,
\ee
where $g(E)$ is an arbitrary $C^\infty$ function, as a power series in the temperature,
\be
I=\int_0^\mu g(E)  \rd E+ \sum_{n=1}^\infty {1\over \beta^{2n}} \left(2 -{1\over 2^{2n-2}}\right) \zeta(2n)g^{(2n-1)}(\mu).
\ee
It is easy to see that this can be written as the operator expansion,
 \be
\label{oncscform}
{1 \over \re^{\beta(\hat H-\mu)}+1}={\pi \partial_{\mu} \over \beta} \csc\left({\pi\partial_{\mu} \over \beta } \right)\Theta(\mu-\hat{H})\ .
\ee

Using now (\ref{trA}), we express the average (\ref{GCvev}), for $\beta=1$, $\eta=-1$, as
\be
\label{somex}
{1\over \Xi} \langle \CO \rangle^{\rm GC}=\tr \left(  {\CO \over \re^{ \hat H-\mu}+1} \right)= \pi \partial_{\mu}\csc(\pi\partial_{\mu}) n_\CO(\mu),
\ee
where
\be
\label{nO}
\ba
n_\CO(\mu)&=
\int\frac{\rd p \rd q}{2\pi\hbar} \Theta(\mu- \hat H)_{\rm W} \CO_{\rm W}\\
&= \int\frac{\rd p \rd q}{2\pi\hbar}
\Theta(\mu- H_{\rm W}) \CO_{\rm W} + \sum_{r=2}^{\infty} {(-1)^r\over r!} \int {\rd q \rd p \over 2 \pi \hbar} \CG_r \delta^{(r-1)}(\mu-H_{\rm W}) \CO_{\rm W}\ .
\ea
\ee
In writing this we have used, in the first line, the fact that the star product drops out of the trace when only two operators are involved \cite{hillery}, and in the second line we used
the Wigner--Kirkwood expansion of the distribution operator. These two expressions, (\ref{somex}) and (\ref{nO}), will be our basic tools to calculate
vevs of Wilson loops in the Fermi gas approach to ABJM theory.

%
%
%
%
%
%

\subsection{Review of the Fermi gas approach}

We will now review the Fermi gas approach to $\CN=3$ Chern--Simons--matter theories, developed in \cite{mp}. We will consider the generalization of ABJM theory given by necklace quivers with $r$ nodes \cite{quiver1,quiver2}, and with fundamental matter in each node. These theories have a gauge group
\be
U(N)_{k_1} \times U(N)_{k_2} \times \cdots U(N)_{k_r}
\ee
and each node will be labelled with the letter $a=1, \cdots, r$. There are bifundamental chiral superfields $A_{a a+1}$, $B_{a a-1}$ connecting
adjacent nodes, and in addition we will suppose that there are $N_{f_a}$ matter superfields $(Q_a, \tilde Q_a)$ in each node, in the fundamental representation. We will write
\be
k_a=n_a k,
\ee
and we will assume that
\be
\label{add0}
\sum_{a=1}^r n_a=0.
\ee

According to the general localization
computation in \cite{kwy}, the matrix model computing the $\IS^3$ partition function of a necklace quiver is given by
\be
\label{quivermm}
Z(N)={1\over (N!)^r} \int  \prod_{a,i} {\rd \lambda_{a,i} \over 2 \pi}  {\exp \left[ {\ri n_a k\over 4 \pi}\lambda_{a,i}^2 \right] \over \left( 2 \cosh{\lambda_{a,i} \over 2}\right)^{N_{f_a}} } \prod_{a=1}^r  {\prod_{i<j} \left[ 2 \sinh \left( {\lambda_{a,i} -\lambda_{a,j} \over 2} \right)\right]^2 \over \prod_{i,j} 2 \cosh \left( {\lambda_{a,i} -\lambda_{a+1,j} \over 2} \right)}.
\ee
The building block of the integrand in (\ref{quivermm}) is the following $N$-dimensional kernel, associated to an edge
connecting the nodes $a$ and $b$:
\be
\label{multiK}
K_{ab}(\lambda_1, \cdots, \lambda_N; \mu_1, \cdots, \mu_N)= {1\over N!} \prod_{i=1}^N \re^{-U_a(\lambda_i)} {\prod_{i<j}  2 \sinh \left( {\lambda_{i} -\lambda_{j} \over 2 k} \right)  2 \sinh \left( {\mu_{i} -\mu_{j} \over 2 k} \right)\over \prod_{i,j} 2 \cosh \left( {\lambda_{i} -\mu_j \over 2 k} \right)}.
\ee
Here,
\be
U_a(\lambda)=-{\ri n_a \over 4 \pi k}\lambda^2 +N_{f_a} \log\left( 2 \cosh{\lambda \over 2 k}\right)
\ee
and will be interpreted as a one-body potential for a Fermi gas with $N$ particles. We denoted by $\lambda_i$ the variables corresponding to the $a$ node,
and by $\mu_i$ those corresponding to the $b$ node, after rescaling them as $\mu, \lambda \rightarrow \mu/k, \lambda/k$.

We now want to interpret the kernel  (\ref{multiK}) as a matrix element
\be
K_{ab}(\lambda_1, \cdots, \lambda_N; \mu_1, \cdots, \mu_N)={1\over N!} \left\{ \lambda_1, \cdots, \lambda_N \right| \hat \rho_{ab} \left| \mu_1, \cdots, \mu_N\right\},
\ee
in terms of a non-symmetrized density matrix $\hat \rho_{ab}$ (i.e. a density matrix for distinguishable particles). We first notice that
\be
\label{symrhoK}
{1\over N!}  \left\{ \lambda_1, \cdots, \lambda_N \right| \hat \rho_{ab} \left| \mu_1, \cdots, \mu_N\right\}={1\over N!} \sum_{\sigma \in S_N} (-1)^{\epsilon(\sigma)}
\rho_{ab} \left( \lambda_1, \cdots, \lambda_N; \mu_{\sigma(1)}, \cdots, \mu_{\sigma(N)}\right).
\ee
We now use the Cauchy identity
 \be
 \label{cauchy}
 \ba
  {\prod_{i<j}  \left[ 2 \sinh \left( {\mu_i -\mu_j \over 2}  \right)\right]
\left[ 2 \sinh \left( {\nu_i -\nu_j   \over 2} \right) \right] \over \prod_{i,j} 2 \cosh \left( {\mu_i -\nu_j \over 2} \right)}
 & ={\rm det}_{ij} \, {1\over 2 \cosh\left( {\mu_i - \nu_j \over 2} \right)}\\
 &=\sum_{\sigma \in S_N} (-1)^{\epsilon(\sigma)} \prod_i {1\over 2 \cosh\left( {\mu_i - \nu_{\sigma(i)} \over 2} \right)}.
 \ea
  \ee
  In this equation, $S_N$ is the permutation group of $N$ elements, and $\epsilon(\sigma)$ is the signature of the permutation $\sigma$.
  We obtain,
\be
\ba
K_{ab}(\lambda_1, \cdots, \lambda_N; \mu_1, \cdots, \mu_N)&={1\over N!} \prod_{i=1}^N \re^{-U_a(\lambda_i)} {\rm det}_{ij} \left( {1\over 2 \cosh {\lambda_i-\mu_j\over 2k}} \right)\\
&=
{1\over N!} \sum_{\sigma \in S_N} (-1)^{\epsilon(\sigma)} \prod_{i=1}^N \re^{-U_a(\lambda_i)} \prod_{i=1}^N t\left({ \lambda_i-\mu_{\sigma(j)} \over k} \right)
\ea
\ee
where we denoted
\be
t(x)={1\over 2 \cosh {x\over 2}}.
\ee
By comparing with (\ref{symrhoK}), it follows that
\be
\rho_{ab}\left( \lambda_1, \cdots, \lambda_N; \mu_1, \cdots, \mu_N\right)=\prod_{i=1}^N \re^{-U_a(\lambda_i)} \prod_{i=1}^N t\left({\lambda_i-\mu_i \over k} \right).
\ee
Since the density matrix is completely factorized, the $N$-particle system is an ideal gas, albeit with a non-trivial one-particle Hamiltonian. By taking the Wigner transform of this expression, with
\be
\hbar =2 \pi k,
\ee
we see that $\rho_{ab}$ defines an $N$-body Hamiltonian
\be
\rho_{ab}^{\rm W} = \re_{\star}^{-H^{ab}_{N,\rm W}},
\ee
where
\be
H^{ab}_{N,\rm W}=\sum_{i=1}^N H^{ab}_{\rm W}(i).
\ee
The one-particle Hamiltonian $H^{ab}_{\rm W}$ is defined by
\be
\re_\star^{-H^{ab}_{\rm W}}= \re^{-U_a(q)} \star \re^{-T(p)}
\ee
and
\be
\label{tp}
T(p)=\log \left( 2 \cosh {p \over 2}\right).
\ee
We can now use repeatedly the resolution of the identity (\ref{resid}) to write the matrix integral (\ref{quivermm}) as
\be
Z(N)=\tr(\hat \rho),
\ee
where $\hat \rho$ is the density matrix
\be
\hat \rho =\hat \rho_{12} \hat \rho_{23} \cdots \hat \rho_{r-1 r} \hat \rho_{r1},
\ee
and this defines the one-particle Hamiltonian $H_{\rm W}$ by
\be
\re_\star^{-H_{\rm W}}=\re_\star^{-H^{12}_{\rm W} } \star \, \re_\star^{-H^{23}_{\rm W}} \star \cdots \star \, \re_\star^{-H^{r-1 r}_{\rm W}} \star \,  \re_\star^{-H^{r1}_{\rm W}}
\ee
For necklace theories without fundamental matter it is easy to see that the total, one-particle Hamiltonian is given by \cite{mp}
\begin{equation}
\re_\star^{-H_{\rm W}}=\frac{1}{2\cosh\frac{p}{2}}\star
\frac{1}{2\cosh\frac{p-n_1q}{2}}\star
\frac{1}{2\cosh\frac{p-(n_1+n_2)q}{2}}\star
\cdots \star
\frac{1}{2\cosh\frac{p-(n_1+\cdots+n_{r-1})q}{2}}.
\end{equation}

\sectiono{Wilson loops in the Fermi gas approach}

\subsection{Incorporating Wilson loops}

We will now show how the calculation of Wilson loops maps to the calculation of statitical-mechanical averages in the Fermi gas approach.
For simplicity, we restrict ourselves to ABJM theory. The general $\CN=3$ quiver can be obtained by a straightforward generalization.

The ABJM quiver is defined by two nodes, with CS levels $k$ and $-k$ (as we mentioned before, and without loss of generality, we will take $k$, the level in the first node, to be positive). The one-body Hamiltonians associated to the edges are given by
\be
\re_{\star}^{-H^{12}_{\rm W}}=\re^{\ri q^2 \over 2 \hbar} \star {1\over 2\cosh{p \over 2} }, \qquad \re_{\star}^{-H^{21}_{\rm W}}=\re^{-{\ri q^2 \over 2 \hbar}} \star {1\over 2\cosh{p \over 2} }.
\ee
Let us consider a $1/6$ BPS Wilson loop with winding number $n$ for the first node. As shown in \cite{kwy} and reviewed above, this corresponds to inserting
\be
\label{wilson-one}
\CO_n=\sum_{i=1}^N \re^{n\lambda_i/k}
\ee
in the matrix integral, after rescaling $\lambda \rightarrow \lambda/k$. The unnormalized vev can be written, in the language of many-body physics, as
\be
\langle \CO_n \rangle ={1\over N!^2} \int \rd \lambda \rd \mu \{ \lambda_1 \cdots \lambda_N | \CO_n \hat \rho_{12} | \mu_1 \cdots \mu_N \} \{ \mu_1 \cdots \mu_N |  \hat \rho_{21} | \lambda_1 \cdots \lambda_N \},
\ee
If we integrate over $\mu$ by using the resolution of the identity, we find
\be
\langle \CO_n\rangle ={1\over N!} \int \rd \lambda  \{ \lambda_1 \cdots \lambda_N | \CO_n \hat \rho_{12} \hat \rho_{21} | \lambda_1 \cdots \lambda_N \} =\tr\left( \CO_n \hat \rho_{12} \hat \rho_{21}\right)
\ee
which is the vev of the one-body operator (\ref{wilson-one}) in an ideal Fermi gas of $N$ particles with one-body Hamiltonian
\be
\re_\star ^{-H_{\rm W}}= {1\over 2 \cosh {p-q \over 2}} \star {1\over 2 \cosh {p\over 2}}.
\ee
Notice that this Hamiltonian is {\it not} Hermitian. This corresponds to the fact that the vev of a $1/6$ BPS operator is not real. After a canonical transformation,
\be
q\rightarrow q+p, \qquad p \rightarrow  p,
\ee
we obtain a more convenient form,
\be
\label{qH}
\re_\star ^{-H_{\rm W}}= {1\over 2 \cosh {q \over 2}} \star {1\over 2 \cosh {p\over 2}}=\re_\star^{-U(q)} \star \re_\star^{-T(p)}
\ee
where $T(p)$ is given in (\ref{tp}) and
\be
\label{uq}
U(q)= \log \left( 2 \cosh {q \over 2} \right).
\ee
After this canonical transformation, the insertion of the BPS Wilson loop corresponds to considering in the Fermi gas a one-body operator of the form
\be
\label{npq}
\CO_n=\exp\left({n(q+p) \over k}\right),
\ee
which we have writen already in the one-particle sector.

Since the vev of a $1/2$ BPS Wilson loop can be obtained from (\ref{sumw}) by considering the vev of the $1/6$ BPS Wilson loop and its conjugate,
we will focus on the analysis of the latter, and deduce the former from (\ref{sumw}).

\subsection{Quantum Hamiltonian and Wigner--Kirkwood corrections}

In the Fermi gas approach of \cite{mp}, the full quantum Hamiltonian contains $\hbar$ corrections and it is not known in closed form. Its semiclassical expansion can be obtained by using the Baker--Campbell--Hausdorff (BCH) formula as applied to the $\star$-product in (\ref{qH}). It was shown in \cite{mp} that, in the calculation of the grand potential of the system,
only the first quantum correction is needed (up to exponentially small corrections in the chemical potential). However, in the calculation of the Wilson loop vev, we will
need an infinite series of terms appearing in the semiclassical expansion, of the form
\be
U^{(n)}(q) \left( T'(p)\right)^n, \qquad T^{(n)}(p) \left( U'(q)\right)^n.
\ee
The coefficients of these terms can be determined in closed form by exploiting some particular
cases of the BCH formula (see \cite{zachos} for examples of such calculations). We will now determine these coefficients.

Let us consider the $\star$ product
\be
\re_\star^A \star \re_\star ^B,
\ee
where
\be
A=-U(q), \qquad B= -a p,
\ee
and $a$ is a constant. The function $U(q)$ is arbitrary. In this case, $B$ acts as the derivative $\ri a \hbar \partial_q$, and the commutator reads
\be
[A, B]_\star =\ri a \hbar U'(q).
\ee
Since this commutator commutes in turn with $B$, one can use a simpler version of BCH which says that (see for example \cite{zachos})
\be
\label{hada}
Z=\log_\star \left( \re_\star^A \star \re_\star ^B\right) = B + A { B]_\star \over 1-\re^{-B]_\star }}
\ee
where $B]_\star $ is to be understood as the operation of performing a $\star$-commutator with $B$ (acting on the left),
and its $n$-th power is obtained by doing the $\star$-commutator $n$ times. The function appearing in
this expression has the well-known expansion
\be
{x \over 1-\re^{-x}} = \sum_{n=0}^{\infty} c_n x^n =1 +{x\over 2} +{x^2 \over 12}+\cdots
\ee
where
\be
c_n={B_n (-1)^n \over n!}
\ee
and $B_n$ are the Bernoulli numbers.
Note that, due to a well-known property of the Bernoulli numbers, all the powers in this series are even except for the second term.
We then conclude that
\be
Z= -a p -\sum_{n\ge 0} c_n (-\ri \hbar a)^n U^{(n)}(q).
\ee

Let us apply this to our case. If we take $T(p)=p/2$, only the terms of the form
\be
U^{(n)}(q) \left( T'(p)\right)^n
\ee
survive in the BCH expansion. This corresponds to choosing $a=1/2$ in the above formula.
We conclude that these terms appear in the Hamiltonian in the form,
\be
\label{lham}
T(p) + U(q)+ \sum_{n\ge 1} {B_n \over n!} (\ri \hbar)^n U^{(n)}(q)\left( T'(p)\right)^n.
\ee
We can calculate in a similar way the terms obtained by exchange of $p$ and $q$.

In our case, where $U(q)$ is given by (\ref{uq}), the derivatives of $U(q)$ can be written in terms
of polylogarithms. Indeed, one finds by direct computation that,
\be
U''(q)= {1\over 4 \cosh^2{q\over 2}}= -{\rm Li}_{-1}(-x),
\ee
where
\be
x=\re^{-q}.
\ee
Therefore, higher derivatives produce polylogarithms of lower order,
\be
U^{(m)}(q)=(-1)^{m+1}{\rm Li}_{1-m}(-x), \qquad m\ge 2.
\ee

\begin{remark} Surprisingly, the real part of the Hamiltonian (\ref{lham}), when $T'(p)=1/2$, can be written in a very suggestive form:
\be
{p +q\over 2} + H_{\rm W}^{\rm q},
\ee
where
\be
H_{\rm W}^{\rm q}=-\sum_{n=0}^{\infty} {B_{2n} \over (2n)!} \left( {\ri \hbar \over 2} \right)^{2n} {\rm Li}_{1-2n}(-x)\ .
\ee
On the other hand, the contribution to the free energy of the resolved conifold with $g\ge1$ is
\be
F(x, g_s)=\sum_{g=1}^{\infty} {B_{2g} \over 2g (2g-2)!} {\rm Li}_{3-2g}(\re^{-t}) g_s^{2g-2}
\ee
It follows that the ``quantum'' Hamiltonian is related to the free energy as
\be
-{\rd \over \rd \tilde{g_s}} \left( {H_{\rm W}^{\rm q} \over \tilde{g_s}} \right)= \left( x {\rd \over \rd x} \right)^2 F(-x, \tilde{g}_s)
\ee
where
\be
\tilde{g}_s ={\ri \hbar \over 2}\ ,
\ee
is identified with the topological string coupling.
\end{remark}

\begin{remark} By using (\ref{hada}) twice, one can compute
\be
\label{ut}
-\log\left( \re^{-U(q)/2} \star \re^{-a p} \star \re^{-U(q)/2} \right)=a p - \sum_{k\ge 0} (\hbar a)^{2k}  {|B_{2k} | \over (2k)! }U^{(2k)}(q).
\ee
This determines the coefficients of $ (T'(p))^{2k} U^{(2k)}(q)$ in the Hermitian Hamiltonian originally considered in \cite{mp}. The result
\be
\label{tu}
-\log\left(\re^{-a q}\star \re^{-T(p)}\star \re^{-aq}\right)=2a q +T(p)+\sum_{k \ge 1} (a\hbar)^{2k} {2 (2^{2k-1}-1)\over (2k)!}|B_{2k}| T^{(2k)}(p)
\ee
does not follow straightforwardly from (\ref{hada}), but it can be derived by using (\ref{hada}) together with symmetry arguments. This determines the
coefficients of $(U'(q))^{2k} T^{(2k)}(p)$ in the Hermitian Hamiltonian in \cite{mp}. Both series of coefficients, (\ref{ut}) and (\ref{tu}), appear as well in the general expansion
of the so-called symmetric BCH formula, and they can be verified up to high order with the results in \cite{casas}. The explicit, analytic expressions (\ref{ut}) and (\ref{tu})
for these coefficients do not seem to have appeared before in the literature.
\end{remark}

We will now compute all the Wigner--Kirkwood corrections for the simplified Hamiltonian considered above, which is obtained from the equation
\be
\label{sH}
\re_\star ^{-H_{\rm W}}= \re^{-U(q)} \star \re^{-a p}.
\ee
More precisely, we want to compute the generating functional of Wigner--Kirkwood corrections obtained by considering the Wigner transform of the
canonical density matrix
\be
\label{wtprop}
\re_\star ^{-t H_{\rm W}},
\ee
as explained in (\ref{genwk}). To calculate (\ref{wtprop}), we use the following trick, inspired by similar calculations in \cite{zachos}.
Let us suppose that we can write (\ref{wtprop}) as
\be
\re_\star^{-t H_{\rm W}}=\re_\star^{-t G(q)} \star \re_\star^{-t a p}
\ee
by using the BCH formula. If this is the case, the $\star$-product can be easily evaluated to obtain,
\be
\label{h-final}
\re_\star^{-t H_{\rm W}}=\re_\star^{-t G(q)} \re^{ {\ri \hbar \over 2} {\overleftarrow{\partial}}_q {\overrightarrow{\partial}}_p  }\, \re_\star^{-t a p}
= \exp\left( -t a p -t \re^{{\xi t \over 2} \partial} G(q)\right).
\ee
where we have denoted
\be
\xi =-\ri a \hbar.
\ee
To obtain the explicit form of $G(q)$, we use the BCH formula (\ref{hada}) to find,
\be
\re_\star^{-t G(q) } \star \re_\star^{-t a p}=\exp_\star \left( - t a p - t \sum_{m\ge 0} c_m (t \xi)^m G^{(m)}(q)\right).
\ee
By construction, this equals $\re_\star^{-t H_W}$. On the other hand, we know that
\be
H_{\rm W}= a p + \sum_{n\ge 0} c_n \xi^n U^{(n)}(q).
\ee
It follows that,
\be
{ \xi t \partial \over 1-\re^{-\xi t \partial} } G(q)= { \xi  \partial \over 1-\re^{-\xi  \partial} } U(q),
\ee
and we find
\be
G(q)={1\over t} {1-\re^{-\xi t \partial} \over 1-\re^{-\xi  \partial}} U(q).
\ee
We conclude that
\be
\label{qprop}
\re_\star^{-t H_{\rm W}}= \exp\left( -t a p - {\re^{{\xi t \over 2} \partial} - \re^{-{\xi t \over 2} \partial} \over 1-\re^{-\xi  \partial}} U(q) \right).
\ee
The second term in the exponent can be computed by using, for example, the definition of Bernoulli polynomials,
\be
{z \re^{-z t} \over 1-\re^{-z}} = \sum_{n\ge 0} B_n (t) (-1)^n {z^n \over n!}.
\ee
One finds,
\be
{1\over t} {\re^{{\xi t \over 2} \partial} - \re^{-{\xi t \over 2} \partial} \over 1-\re^{-\xi  \partial}} U(q) = \sum_{m\ge 0} {B_{m+1}(t/2)-B_{m+1}(-t/2) \over t} {(-1)^m \over (m+1)!} \xi^m U^{(m)}(q).
\ee
Since $\CG_0=1$, $\CG_1=0$, the exponent in (\ref{qprop}) should be of the form
\be
-t H_{\rm W} +\CO(t^2).
\ee
Indeed, by using
\be
{B_{m+1}(t/2)-B_{m+1}(-t/2) \over m+1 }= B_m t +\CO(t^3)
\ee
one sees that
\be
ap+{1\over t} {\re^{{\xi t \over 2} \partial} - \re^{-{\xi t \over 2} \partial} \over 1-\re^{-\xi  \partial}} U(q) =H_{\rm W}(q,p) +\CO(t^2).
\ee
The expression (\ref{qprop}) generates all the functions $\CG_r$ by expanding in $t$, and one can verify the results, at the very first orders, against the explicit expression in terms of $\star$-products given in (\ref{Gr}).

We also notice that the operator appearing in (\ref{qprop}) can be written as
\be
\label{shift1}
 {\re^{{\xi t \over 2} \partial} - \re^{-{\xi t \over 2} \partial} \over 1-\re^{-\xi  \partial}} U(q) =  {1\over 1-\re^{-\xi  \partial}} \left[ U\left( q+{\xi t \over 2} \right)
-U\left( q-{\xi t \over 2} \right) \right].
\ee
One has to be however careful, since the expansion of the differential operator in the denominator leads to
\be
\label{shift2}
 \sum_{\ell \ge 0} {B_\ell (-1)^\ell \over \ell!} \xi^{\ell-1} \partial^{\ell-1}  \left[ U\left( q+{\xi t \over 2} \right)
-U\left( q-{\xi t \over 2} \right) \right],
\ee
and one has to define properly the term $\ell=0$, which involves $\partial^{-1}$. Comparison with the expansion in terms of Bernoulli numbers shows that this term stands for,
\be
\label{n0ex}
{1\over \xi } \partial^{-1}  \left[ U\left( q+{\xi t \over 2} \right)
-U\left( q-{\xi t \over 2} \right) \right] = t \sum_{g=0}^\infty {1\over (2g+1)!} \left( {t \xi \over 2} \right)^{2g} U^{(2g)}(q).
\ee
This can be also written in terms of an integral,
\be
\label{n0int}
\ba
& \sum_{g=0}^\infty {1\over (2g+1)!} \left( {t \xi \over 2} \right)^{2g} U^{(2g)}(q)\\
&= {1\over t \xi} \int_{\Lambda}^q \left[ U\left( q+{\xi t \over 2} \right)
-U\left( q-{\xi t \over 2} \right) \right] +  \sum_{g=0}^\infty {1\over (2g+1)!} \left( {t \xi \over 2} \right)^{2g} U^{(2g)}(\Lambda),
\ea
\ee
where $\Lambda$ is an appropriate reference point.

As we will see in the next subsection, we need the expression of the canonical density matrix for the value
\be
\label{tn}
t={2n \over k},
\ee
where $n$ is the winding of the Wilson loop operator.
This can be evaluated in principle with (\ref{qprop}) and (\ref{shift2}). However, the calculation is rather delicate, since the shift by
\be
\label{shift}
\xi t/2=-n \pi\ri
\ee
implies that we are resumming the series of semiclassical corrections beyond its radius of convergence, and a regularization is needed. We will proceed as follows.
First, we notice that, in the ``polygonal'' limit $|q| \rightarrow \infty$,
\be
U(q) \approx {|q| \over 2} + \CO\left(\re^{-|q|}\right),
\ee
In this limit,  the second term in the exponent of (\ref{qprop}) is given, for $q\not=0$, by
\be
\label{poly-limit}
-t |q| -{t \xi \over 2}  {\rm sgn}(q)
\ee
since only $U(q)$ and its first derivative survive. Therefore, we want to calculate the correction to this polygonal limit for the value of $t$ given by (\ref{tn}). In fact, this  correction is given by a distribution supported at $q=0$, which can be obtained as follows. The first term of (\ref{shift2})
can be computed by writing, for $q>0$,
\be
\label{tU}
U(q)={q\over 2} +\widetilde U(q), \qquad \widetilde U(q)=\log (1+ \re^{-q}).
\ee
We have to calculate the sum appearing in the r.h.s. of (\ref{n0ex}), which we write as
\be
{q\over 2} + \sum_{g=0}^\infty {1\over (2g+1)!} \left( {t \xi \over 2} \right)^{2g} \widetilde U^{(2g)}(q).
\ee
Since this function, as well as all its derivatives, vanish at infinity, we take $\Lambda=\infty$ as a reference point. For the particular value (\ref{shift}), we obtain from (\ref{n0int})
\be
\sum_{g=0}^\infty {1\over (2g+1)!} \left( {t \xi \over 2} \right)^{2g} \widetilde U^{(2g)}(q)= {1\over t\xi} \int_{\infty}^q \left[ \widetilde U\left( q+{\xi t \over 2} \right)
-\widetilde U\left( q-{\xi t \over 2} \right) \right] =0
\ee
since the integrand vanishes. We conclude that the term with $\ell=0$ in (\ref{shift2}) is given by
\be
{t\over 2}q, \qquad q>0.
\ee
A similar reasoning for $q<0$ shows that the first term in (\ref{shift2}), for the value (\ref{tn}) of $t$, is
\be
-{n \pi \ri \over \xi} |q|.
\ee
The second term in (\ref{shift2}) involves the derivative of this first term. Equivalently, it can be computed as the monodromy of $U(q)$,
\be
 U\left( q+{\xi t \over 2} \right)
-U\left( q-{\xi t \over 2} \right) =  -n  \pi \ri \, {\rm sgn}(q).
\ee
We then see from (\ref{shift2}) that the full series of corrections involves the distribution $\CS(q)$ defined as
\be
\label{csq}
\CS(q)= \sum_{\ell \ge 0} {B_\ell (-1)^\ell \over \ell!} \xi^{\ell-1} \partial^{\ell-1} \, {\rm sgn}(q)={1\over 1-\re^{-\xi \partial} } {\rm sgn} (q)= {|q| \over \xi} + {1\over 2} {\rm sgn}(q) + \CO(\xi).
\ee
To calculate $\CS(q)$, we take a derivative w.r.t. $q$, and we multiply both sides by $1 -\re^{-\xi \partial}$. We obtain the equation
\be
\label{teq}
\CT(q)-\CT(q-\xi)= \delta(q) + \delta(-q),
\ee
where
\be
\CT(q)=\CS'(q).
\ee
The Fourier transform of (\ref{teq}) gives
\be
\widehat \CT(\omega)={\sqrt{2 \over \pi}} {1\over 1-\re^{\ri \xi \omega}}.
\ee
We now set
\be
\xi=-\ri \vartheta, \qquad \vartheta=\pi k,
\ee
and solve for $\CT(q)$ by doing an inverse Fourier transform. This transform is in principle ill-defined due to the pole at $\omega=0$, but we can regularize it
in a standard way by taking a principal value at the origin (or an extra derivative w.r.t. $q$). We obtain in this way
\be
\CT(q)=-{{\rm P} \over 2 \pi} \int \rd \omega\, \re^{-\ri \omega q} { \re^{-\omega \vartheta/2} \over \sinh\left( {\omega \vartheta \over 2} \right)}= {\ri \over \vartheta}
\coth\left( {\pi q \over \vartheta} \right).
\ee
We now integrate w.r.t. $q$ to obtain $\CS(q)$. The result is, after fixing the appropriate value for the integration constant,
\be
\CS(q)={1\over 2} + {\ri \over \pi} \log \left( 2  \sinh \left( {\pi q \over \vartheta} \right) \right).
\ee
To see that this is a natural regularization, and to fix the integration constant, we note that for $q>0$ this can be written as
\be
\CS(q)={q\over \xi} +{1\over 2} + {\ri \over \pi} \log \left( 1 -\re^{-2 \pi q/\vartheta}\right),
\ee
while for $q<0$ we find,
\be
\CS(q)=-{q\over \xi} +{1\over 2} + {\ri \over \pi} \log(-1) + {\ri \over \pi} \log \left( 1 -\re^{2 \pi q/\vartheta}\right),
\ee
i.e.
\be
\CS(q)= {|q| \over \xi} + {1\over 2} {\rm sgn}(q) + {\ri \over \pi} \log \left( 1 -\re^{2 \pi |q|/\vartheta}\right),
\ee
so that, for $q\not=0$, and $\xi$ small, we find,
\be
\CS(q)\approx {|q| \over \xi} + {1\over 2} {\rm sgn}(q),
\ee
which is consistent with (\ref{csq}) and also gives the polygonal limit we need, cf. (\ref{poly-limit}).

It might be surprising that an infinite sum of distributions (\ref{csq}) can be resummed to a smooth function of $q$. However, this is standard in the context of the
semiclassical approximation of Wigner functions, and is performed by means of Fourier transforms, as we have just done \cite{voros}. For example, the ground state of a harmonic oscillator in the Wigner formulation involves the Gaussian
\be
\label{wigner-gauss}
f_{\rm W}(q,p)=f(q)f(p), \qquad f(q)={1\over {\sqrt{\pi \hbar} } } \re^{-q^2/\hbar}.
\ee
But
\be
\hat f(\omega)={1\over {\sqrt{2 \pi}}} \re^{-\hbar \omega^2/4}= {1\over {\sqrt{2 \pi}}} \sum_{\ell=0}^{\infty} {(-1)^\ell \hbar^\ell \over 4^\ell \ell!} \omega^{2\ell}
\ee
which has inverse Fourier transform
\be
\sum_{\ell=0}^{\infty} {(-1)^\ell \hbar^\ell \over 4^\ell \ell!} \delta^{2\ell}(q).
\ee
In the classical limit $\hbar \rightarrow 0$ we have a localized particle at the origin, and the $\hbar$ corrections give an infinite sum of distributions which
however can be obtained from the smooth Gaussian in (\ref{wigner-gauss}).

We conclude that, for the simplified Hamiltonian (\ref{sH}), the canonical density matrix with $t$ given in (\ref{tn}) is
\be
\label{exq}
\re_\star ^{ -{2n\over k} H_{\rm W} }=\exp\left[ -{n \over k} p + {n \pi \ri \over 2} - n \log \left(  2 \sinh \left( {q \over k} \right) \right) \right],
\ee
at least with the natural regularization procedure explained above.

\subsection{Integrating over the Fermi surface}

\begin{figure}
\center
\includegraphics[height=8cm]{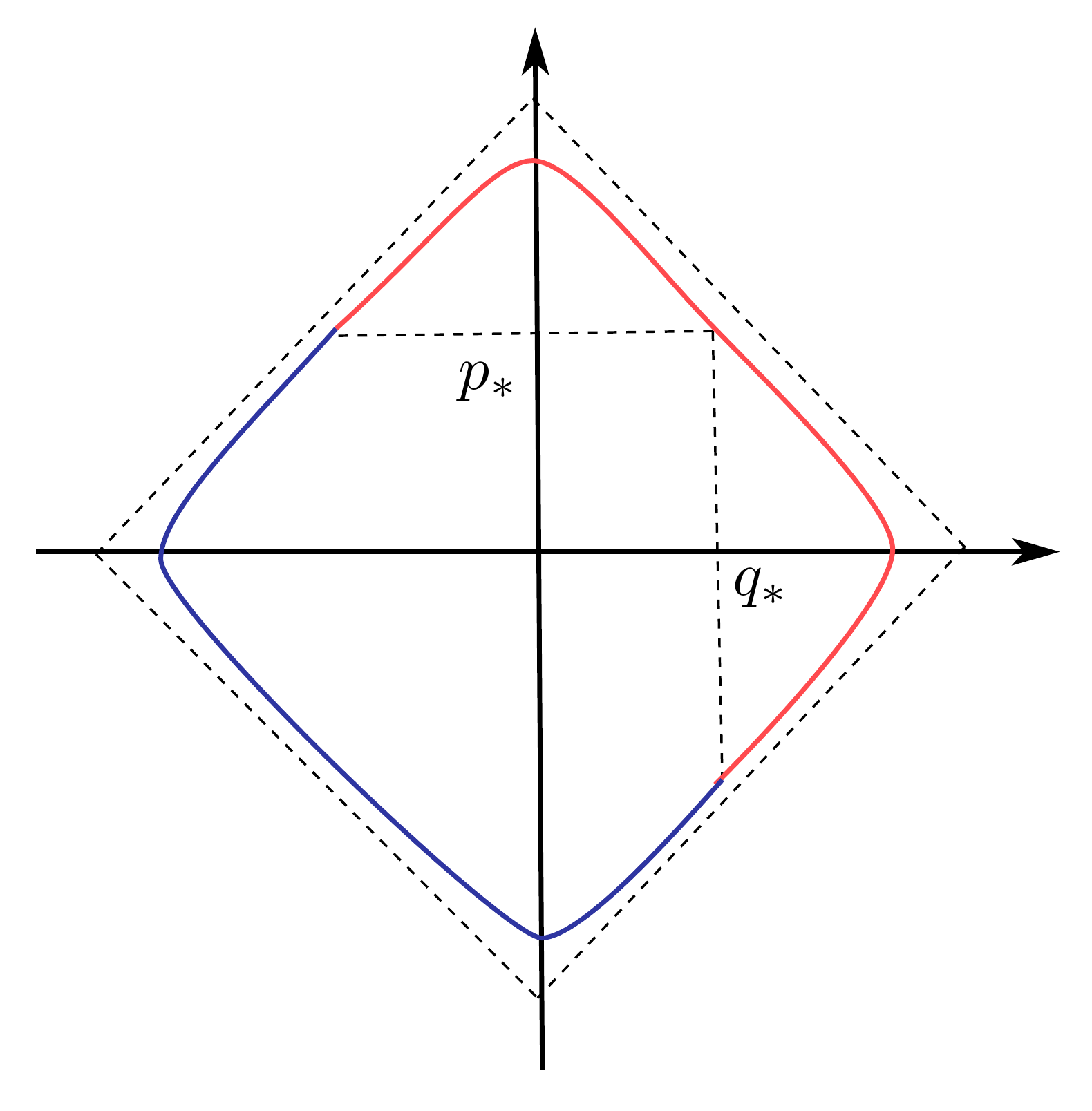}
\caption{The regions in the Fermi surface.}
\label{regions-Wilson}
\end{figure}

We are now ready to calculate the vev of the $1/6$ BPS Wilson loop with
winding number $n$ in the Fermi gas approach. The corresponding one-body
operator is given in (\ref{npq}). The first step is then to calculate
(\ref{nO}) for this operator, i.e.
\be
\label{non}
n_{\CO_n}(\mu)= \int {\rd q \rd p \over 2 \pi \hbar}  \Theta(\mu-H_{\rm W}) \re^{n(q+ p) \over k}+ \sum_{r \ge 1} {(-1)^r \over r!} {\rd^{r-1} \over \rd \mu^{r-1}} \int {\rd q \rd p \over 2 \pi \hbar} \delta(\mu-H_{\rm W}) \CG_r
\re^{n(q+ p)\over k}.
\ee
Notice that, since the Hamiltonian is complex, the Fermi surface
\be
\label{fermisurface}
H_{\rm W}(q,p)=\mu
\ee
is in principle a surface in complex space. However, up to exponentially small corrections,
we can recover a real Hamiltonian by a Wick rotation of the Planck constant, $\hbar \rightarrow -\ri \hbar$, so that $\ri\hbar$ is real.
After doing this, the integration process is perfectly well defined, and we can rotate back at the end of the calculation. Equivalently, it can be easily seen from our computations that this involves integrating
over appropriate paths in the complexified phase space.

The first integral is over the region enclosed by the Fermi surface. However,
by integrating w.r.t. $p$ or $q$ one can reduce the integral to a boundary
integral over the Fermi surface, plus a ``bulk'' contribution which is
easy to calculate. As in \cite{mp}, we will divide the boundary of the Fermi surface in appropriate
regions. The quantum Hamiltonian reads,
\be
H_{\rm W}=T(p) + U(q)+ {\ri \hbar \over 4} U'(q) T'(p)+ \cdots
\ee
where the corrections are exponentially small. The point in the Fermi surface with $p$ coordinate
\be
p_*=\mu +{\ri \hbar \over 8}
\ee
has a $q$ coordinate of the form
\be
q_*= \mu + {\ri \hbar \over 8}+ \CO(\re^{-\mu}).
\ee
It is easy to see that the leading contribution to the Wilson loop is obtained by subtracting the contribution of the bulk region
\be
-p_*\le p \le p_*, \qquad -q_*\le q \le q_*
\ee
to the contribution of the boundary (of course, in writing this inequalities, we assume that we have performed a Wick rotation and that $\ri \hbar$ is real).
But, if we restrict ourselves to terms which are proportional to $\exp (2 n \mu/k)$, the only contribution comes from
the boundary shown in red in \figref{regions-Wilson}. This region can be divided in turn in two regions: a region where
\be
\label{firstregion}
p> p_*, \qquad  -q_*\le q \le q_*,
\ee
and the region obtained by exchanging $p$ and $q$,
\be
q> q_*, \qquad  -p_*\le p \le p_*.
\ee
They give the same contribution, so we will restrict ourselves to the first region and then multiply the result by two.
Along the curve bounding the region (\ref{firstregion}) we can neglect exponentially small terms in $p$, i.e. we can assume that $T(p) =p/2$. We can then write
\be
p(\mu,q)=2\mu + \left( 2 H_{\rm W}- p\right),
\ee
where
\be
2 H_{\rm W}- p= U(q) + {\ri \hbar \over 4} U'(q)+\cdots
\ee
only depends on $q$ and it has been computed in (\ref{lham}), with $T(p)=p/2$. We want to calculate the first term in (\ref{non}),
\begin{equation}
\int {\rd q \rd p \over 2 \pi \hbar} \re^{n(q+ p) \over k} \Theta(\mu-H_{\rm W})={k \over n} \int {\rd q  \over 2 \pi \hbar} \re^{q\over k} \left( \re^{n p(\mu,q) \over k}-1\right)
\ee
and we restrict to terms which are proportional to $\exp (2 n \mu/k)$, so we keep only the first term. After plugging in the value of $p(\mu,q)$, we find
\be
\label{ints}
{k \over 2 \pi n \hbar} \re^{\frac{2 n \mu}{k}}  \int_{-q_*}^{q_*} \rd q \, \re^{ n(p+q) \over 2k} \re^{ -{2n \over k} H_{\rm W} }.
\end{equation}
Notice that the $p$ dependence in this and similar expressions cancels trivially.
It is easy to see that all $\hbar $ corrections to the Hamiltonian contribute to this integral, even if we neglect exponentially small corrections.

Let us now consider the Wigner--Kirkwood corrections to (\ref{non}) along the curve bounding the region (\ref{firstregion}). By writing
\be
 \delta(\mu-H_{\rm W}(q,p))={1\over \left| {\partial H_{\rm W}(q,p) \over \partial p} \right|} \delta(p-p(\mu,q))=2 \delta(p-p(\mu,q))
 \ee
 we obtain
\be
\ba
&2 \sum_{r \ge 1} {(-1)^r \over r!} {\rd^{r-1} \over \rd \mu^{r-1}} \re^{2 n \mu \over k}\int_{-q_*}^{q_*}{\rd q \over 2 \pi \hbar} \CG_r
\re^{n(q+ p) \over k } \re^{ -{2n \over k} H_{\rm W} } \\
&={k  \over n} \re^{2 n \mu \over k} \sum_{r \ge 1} {(-1)^r \over r!} \left( {2 n \over k}\right)^{r} \int_{-q_*}^{q_*}{\rd q \over 2 \pi \hbar} \CG_r
\re^{n(q+ p) \over k } \re^{ -{2n \over k} H_{\rm W} }.
\ea
\ee
This combines with (\ref{ints}) to produce,
\be
{k \over 2 \pi n \hbar }  \re^{2 n \mu \over k} \sum_{r \ge 0} {(-1)^r \over r!} \left( {2 n \over k}\right)^{r} \int_{-q_*}^{q_*}\rd q \,  \CG_r
\re^{n(q+ p) \over k } \re^{ -{2n \over k} H_{\rm W} }= {k \over 2 \pi n \hbar}  \re^{2 n \mu \over k} \int_{-q_*}^{q_*}\rd q \, \re^{n(q+ p) \over k }  \re_\star ^{ -{2n\over k} H_{\rm W} }.
\ee
Using (\ref{exq}) we find that this integral equals
\be
\label{inte}
{ k \over 2 \pi n \hbar} \ri ^n \re^{\frac{2 n \mu}{k}}  \int_{-q_*}^{q_*} \rd q {\re^{n q/k} \over \left( 2 \sinh \left( {q \over k} \right) \right)^n },
\ee
where
\be
q_*\approx \mu +{\pi \ri k \over 4}.
\ee
There is a singularity of the integrand for $q=0$. However, as we will
see, this can be avoided in a natural way.
Also notice that, as $q_*\rightarrow \infty$, the integral diverges due to the upper integration limit, but it is not divergent when we send the lower integration limit
to infinity. In fact, doing this only introduces exponentially small corrections (which we are neglecting anyway), and up to these corrections we can just compute,
\be
k \, I_n= \int_{-\infty - {\pi \ri k \over 4}}^{q_*} \rd q {\re^{n q/k} \over \left( 2 \sinh \left( {q \over k} \right) \right)^n }.
\ee
To calculate this integral, we make the following change of variables
\be
u=\re^{q/k},
\ee
so that
\be
I_n= \int_{0}^{u_*} \rd u {u^{n-1} \over (u-u^{-1})^n }.
\ee
It is actually simpler to calculate the generating functional,
\be
\CI=\sum_{n=1}^\infty I_n z^n=\int_{0}^{u_*} \rd u {z u \over u^2 (1-z)-1 }\approx {1\over 2} {z\over 1-z} \log (-u_*^2) +{1\over 2} {z\over 1-z} \log (1-z),
\ee
where
\be
-u_*^2 = \re^{2 \mu/k} \re^{-{\ri \pi \over 2}}
\ee
and we have again neglected exponentially small corrections.
We now take into account that
\be
-{\log (1-z) \over 1-z} = \sum_{n=1}^\infty H_n z^n,
\ee
where $H_n$ are harmonic numbers, to obtain
\be
k I_n= \mu-{\ri \pi k \over 4} -{k \over 2} H_{n-1}.
\ee
Notice that the integrand above has poles at $u^2=(1-z)^{-1}$, and we have chosen an integration contour in the complex $u$-plane
which avoids these poles. This is natural since the upper limit
of integration, $u_*$, is in fact complex.

Putting all together, we obtain
\be
{k  \over 2 \pi n \hbar} \re^{\frac{2 n \mu}{k}}\ri^n \left( \mu -{\pi \ri k  \over 4}  -{k\over 2} H_{n-1} \right).
\ee
As we explained above, there is an identical contribution from the region obtained by exchanging $p \leftrightarrow q$. Finally, one has to subtract the contribution from the bulk region, which gives
\be
-\int_{-q_*}^{q_*} \int_{-p_*}^{p_*} {\rd q \rd p \over 2 \pi \hbar}  \re^{n(p+ q)\over k} =-{k^2 \over 2 \pi n^2  \hbar} \re^{ {2 n \mu \over k}+{\ri n \hbar \over 4 k }}+ \cdots =- {\ri^n k^2 \over 2 \pi n^2 \hbar}\re^ {2 n \mu \over k}  +\cdots
\ee
where the dots denote subleading exponentially small corrections. Therefore, up to these corrections, we find
\be
\label{ncon}
n_{\CO_n}(\mu)\approx {k  \over 2 \pi n \hbar} \re^{\frac{2 n \mu}{k}}\ri^n \left( 2 \mu -{\pi \ri k  \over 2}  -k  H_n \right).
\ee
As we will see in a moment, this is in precise agreement with the result obtained in the 't Hooft expansion at genus zero.

According to (\ref{somex}), in order to find the full statistical-mechanical average, we just have to take into account the finite temperature
corrections encoded in the Sommerfeld expansion. We then find,
\begin{eqnarray}
\frac{1}{\Xi}\langle \CO_n \rangle^{\mbox{GC}}=\pi \partial_{\mu}\csc(\pi\partial_{\mu}) n_{\CO_n}(\mu),
\end{eqnarray}
with the value obtained in (\ref{ncon}), which we will write as
\be
n_{\CO_n}(\mu)\approx \big(A(k)\mu+B(k)\big)
\re^{\frac{2n \mu}{k}}\ .
\ee
Here $A(k)$ and $B(k)$ are given by
\begin{eqnarray}
A(k)= { \ri^n \over 2 \pi^2 n},\qquad B(k)=-\frac{k}{4\pi^{2} n}\ri^{n+1} \Big(\frac{\pi}{2} -\ri H_n \Big)\ .
\end{eqnarray}
Putting things together, we find
\begin{eqnarray}\label{Sommerfeld1}
\frac{1}{\Xi}\langle{\cal O}_{n}\rangle^{\mbox{GC}}=\frac{2\pi n}{k}\csc\frac{2\pi n}{k}\Big[
\big(\mu+\frac{k}{2n}-\pi\cot\frac{2\pi n}{k}\big)A(k)+B(k)\Big]\re^{\frac{2n}{k}\mu}\ ,
\end{eqnarray}
where $\Xi$ is the grand-canonical partition function calculated in \cite{mp}, which is given by
\begin{eqnarray}\label{grandpot}
\Xi=\exp\Big(\frac{2\mu^{3}}{3\pi^{2}k}+\frac{\mu}{3k}+\frac{\mu k}{24}\Big),
\end{eqnarray}
up to exponentially small corrections and an overall, $\mu$-independent constant. (\ref{Sommerfeld1}) gives then
the exact grand canonical correlator at all $k$, up to exponentially small corrections in $\mu$. \
To get the original normalized Wilson loop correlator, we have to perform the inverse transformation
\begin{eqnarray}
\label{ai-w}
\langle W_{n}^{1/6}\rangle=\frac{1}{2\pi\ri Z}\int \rd\mu\, \re^{-\mu N}\langle{\cal O}_{n}\rangle^{\mbox{GC}}\ ,
\end{eqnarray}
where $Z$ denotes the partition function of the theory, which is itself given by
\be
\label{four-t}
Z(N)=\frac{1}{2\pi\ri }\int \rd\mu\, \re^{-\mu N} \Xi(\mu).
\ee
We recall that the Airy function has the integral representation
\be
{\rm Ai}(z)={1\over 2 \pi \ri} \int_{\CC} \rd t\, \exp\left( {t^3 \over 3} -z t\right),
\ee
where $\CC$ is a contour in the complex plane from $\re^{-\ri \pi/3}\infty$ to $\re^{\ri \pi/3}\infty$.
Therefore,
\be
\label{ai-z}
Z(N) \propto {\rm Ai}\Big[C^{-1/3}\Big(N-\frac{k}{24}
-\frac{1}{3k}\Big)\Big],
\ee
which is a result first derived in \cite{fhm} for ABJM theory and then rederived in the Fermi gas approach in \cite{mp} for a
class of $\CN=3$ theories. Now, due to the exponential form of (\ref{Sommerfeld1}),
the integral (\ref{ai-w}) can be written in terms of quotients of Airy functions,
\be
\ba\label{airy}
\langle W_{n}^{1/6} \rangle=-&C^{-1/3}A_{1}(k)\frac{
\mbox{Ai}'\Big[C^{-1/3}\Big(N-\frac{k}{24}
-\frac{6n+1}{3k}\Big)\Big]}{\mbox{Ai}\Big[C^{-1/3}\Big(N-\frac{k}{24}
-\frac{1}{3k}\Big)\Big]}\\
& +A_{2}(k)\frac{\mbox{Ai}\Big[C^{-1/3}\Big(N-\frac{k}{24}
-\frac{6n+1}{3k}\Big)\Big]}{\mbox{Ai}\Big[C^{-1/3}\Big(N-\frac{k}{24}
-\frac{1}{3k}\Big)\Big]}\ ,
\ea
\ee
where the prime denotes the derivative of the Airy function, and
\be
C=\frac{2}{\pi^{2}k}.
\ee
The functions $A_{1}(k)$ and $A_{2}(k)$ are defined as
\begin{eqnarray}\label{A1A2}
&&A_{1}(k)=\frac{2\pi n}{k}\csc\frac{2\pi n}{k}\, A(k)\ ,\\
&&A_{2}(k)=\frac{2\pi n}{k}\csc\frac{2\pi n}{k}\Big[\big(\frac{k}{2n}-\pi\cot
\frac{2\pi n}{k}\big)A(k)+B(k)\Big]\ .
\end{eqnarray}
Once the answer for the 1/6 BPS Wilson loop correlator is found, we can obtain the expectation value of the 1/2 BPS Wilson loop via (\ref{sumw})
\be
\langle W_{n}^{1/2} \rangle=\frac{1}{4}\csc\frac{2\pi n}{k}\,\frac{\mbox{Ai}\Big[C^{-1/3}\Big(N-\frac{k}{24}
-\frac{6n+1}{3k}\Big)\Big]}{\mbox{Ai}\Big[C^{-1/3}\Big(N-\frac{k}{24}
-\frac{1}{3k}\Big)\Big]}\ . \label{airy12}
\ee
Notice that the Airy functions in the denominators of (\ref{airy}) and (\ref{airy12}) come from the partition function \cite{fhm,mp}.

We should emphasize that (\ref{airy}) and (\ref{airy12}) are exact results at all orders in the $1/N$ expansion, up to exponentially small corrections. We will now
extract from it some results on the 't Hooft genus expansion and test it with known results at low genus.

\subsection{Genus expansion}

In order to extract the 't Hooft expansion of the Wilson loop correlator, we have to expand (\ref{airy}) in powers of $1/k$. Since we are working in the $1/N$ expansion and
$k$ is generic, the results we will obtain are valid in the strong 't Hooft coupling regime
\be
\label{strongcoup}
\lambda \gg 1\ .
\ee
The 't Hooft expansion of the Wilson loop vevs from the ABJM matrix model has been reviewed and extended in section 2. Therefore, we can compare the genus expansions we obtain from (\ref{airy}) with known results. We will do the expansions explicitly for genus zero, genus one, and genus two. In appendix B, we will summarize the results for few more higher genus expansions.
\vspace{0.5cm}

{\bf\textit{--- Genus zero}}

\vspace{0.3cm}
To test the agreement between the results of ABJM matrix model and the Fermi gas approach, it will be more convenient to expand the Airy functions in (\ref{airy}) in terms of the $\kappa$-variable (\ref{kappalambda}) where only positive powers of $\kappa$ are relevant at strong coupling. For genus zero we find,
\begin{eqnarray}\label{kappa0}
g_{s}^{-1}\langle W_{n}^{1/6}\rangle_{g=0}=\frac{\ri^{n}\kappa^{n}k}{4\pi^{2}n}
\Big(\log\kappa-\frac{\ri\pi}{2}- H_{n}\Big)\ .
\end{eqnarray}
This agrees with the result (\ref{W16kappa}) obtained with standard matrix model techniques.

The strong coupling expansion of this result can be obtained by expanding the Airy functions in (\ref{airy}) in terms of the 't~Hooft coupling $\lambda$ in the regime (\ref{strongcoup}),
\begin{eqnarray}
\label{genus0}
\langle W_{n}^{1/6}\rangle_{g=0}&=&2\pi\ri^{n+1}\Bigg(\frac{\sqrt{\lambda}}{2\sqrt{2}\pi n}-\Big(\frac{H_{n}}{4\pi^{2}n}+
\frac{\ri}{8\pi n}+\frac{1}{96}\Big)
+\Big(\frac{\ri}{192}+\frac{\pi n}{4608}+\frac{H_{n-1}}{96\pi}\Big)\frac{1}{\sqrt{2\lambda}}\nonumber\\
&&-\Big(\frac{\ri\pi n}{18432}+\frac{\pi^{2}n^{2}}{663552}+\frac{nH_{n-1}}{9216}\Big)\frac{1}{\lambda}
+\CO(\lambda^{-3/2})\Bigg)\re^{\pi n\sqrt{2\lambda}}\ .
\end{eqnarray}
Once we have obtained the result for the expectation value of the 1/6 BPS Wilson loop, the result for the 1/2 BPS Wilson loop follows from (\ref{sumw}) by incorporating the result of the other node of the ABJM quiver gauge theory.
\vspace{0.5cm}

{\bf\textit{--- Genus one}}

\vspace{0.3cm}
As the next step of our checks, we would like to compare the results of the Fermi gas approach and the ABJM matrix model at genus one. 
Expanding (\ref{airy}) in terms of $\kappa$, we find the following expression
\be
\ba
\label{genus1kappa}
\langle W_{n}^{1/6}\rangle_{g=1}=-\ri^{n+1}\kappa^{n}  \Bigg[& \frac{n\log \kappa }{12\pi} -\frac{\ri n}{24}-\frac{2nH_{n}+3n-3}{24\pi}\\
& + \left(\frac{3n+1}{24\log \kappa} -\frac{1}{8\log^2 \kappa}\right) \left(\frac{\ri}{2}+\frac{H_{n-1}}{\pi}\right) \Bigg]\ .
\ea
\ee
Using  (\ref{sumw}) we obtain for the vev of the 1/2 BPS Wilson loop at genus one,
\be
\label{12gone}
\langle W_{n}^{1/2}\rangle_{g=1}=-\ri^n \kappa^{n}\Bigg[
\frac{2n\log^{2}\kappa-(3n+1)\log\kappa+3}{24\log^{2}\kappa}\Bigg]\ .
\ee
The genus one result for the 1/6 Wilson loop correlator with winding one was first found in \cite{dmp} by analyzing the ABJM matrix model. If we set $n=1$ in (\ref{genus1kappa}), we find precise agreement between our results and those of \cite{dmp}. We can also easily compute the genus one, 1/2 BPS Wilson loop correlator with arbitrary winding from the
ABJM matrix model, using (\ref{laurantW1}), and the result is in agreement with the general expression (\ref{12gone}).

Expanding the Airy functions in (\ref{airy}) directly in terms of the 't~Hooft coupling $\lambda$ in the region (\ref{strongcoup}), 
we find the following expansion for the 1/6 BPS Wilson loop expectation value
\begin{eqnarray}\label{genus1}
\langle W_{n}^{1/6}\rangle_{g=1}&=&-\frac{\ri^{n+1}\lambda}{2\pi}\Bigg(\frac{\pi n}{3\sqrt{2\lambda}}-\Big(\frac{2n H_{n}+3n-3}
{12}+\frac{\ri\pi n}{12}+\frac{\pi^{2}n^{2}}{144}\Big)\frac{1}{\lambda}+\Big(\frac{(3n+1)\ri}
{24}+
\frac{\ri\pi^{2}n^{2}}{288}+
\frac{\pi^{3}n^{3}}{6912}\nonumber\\
&&+\frac{n(n-1)\pi}{96}+\frac{(3n+1)H_{n-1}}{12\pi}+\frac{\pi n^{2}H_{n-1}}{144}\Big)\frac{1}{\lambda\sqrt{2\lambda}}
-\Big(\frac{\ri}{16\pi}+\frac{\ri\pi(3n+1)n}{1152}\nonumber\\
&&+\frac{\pi^{2}n^{2}(n-1)}{9216}+\frac{\ri\pi^{3}n^{3}}{27648}
+\frac{\ri\pi^{4}n^{4}}{995328}+\frac{(3n+1)n H_{n-1}}{576}+\frac{H_{n-1}}{8\pi^{2}}+\frac{\pi^{2}n^{3}H_{n-1}}{13824}\Big)
\frac{1}{\lambda^{2}}\nonumber\\
&&+\CO(\lambda^{-5/2})
\Bigg)\re^{\pi n\sqrt{2\lambda}}\ .
\end{eqnarray}
Similar to the genus zero result, the expansion of the 1/2 BPS Wilson loop correlator with winding $n$ is automatically obtained by applying (\ref{sumw}).
\vspace{0.5cm}

{\bf\textit{--- Genus two}}

\vspace{0.3cm}
As our last check, we consider ABJM Wilson loop correlators at genus two. Expanding (\ref{airy}) in terms of $\kappa$, we have
\begin{align}
\label{genus2}
g_{s}^{-1}\langle W_{n}^{1/6}\rangle_{g=2} & =-\frac{\ri^n \kappa^n k}{2 \pi n}\left[
 -\frac{7n^4 \log\kappa }{720 \pi } + \frac{7\ri n^4}{1440}-\frac{23n^3}{720 \pi }+\frac{n^4}{48 \pi }+\frac{7n^4 H_n}{720 \pi } \right. \nonumber\\
& \phantom{{} = {} } + \frac{n^2}{\log \kappa } \left( -\frac{\ri n(3n+1)}{288}+\frac{11}{576\pi }+\frac{5 n}{96 \pi }-\frac{n^2}{64 \pi }-\frac{n(3n+1) H_n}{144 \pi } \right) \nonumber\\
& \phantom{{} = {} } + \frac{n^2}{\log^2 \kappa} \left(   \frac{\ri }{1152}+\frac{\ri n}{64}+\frac{\ri n^2}{128}-\frac{7}{96 \pi }-\frac{n}{96 \pi }  +\frac{ H_{n-1}}{576 \pi }+\frac{n H_n}{32 \pi }+\frac{n^2 H_n}{64 \pi }  \right) \nonumber\\
& \phantom{{} = {} } + \frac{n}{\log^3 \kappa} \left(  -\frac{\ri}{1152}-\frac{\ri n}{96}-\frac{5 \ri n^2}{192}+\frac{5 n}{64 \pi }-\frac{H_{n-1}}{576 \pi }-\frac{n H_{n-1}}{48 \pi }-\frac{5 n^2 H_n}{96 \pi } \right) \nonumber\\
&  \phantom{{} = {} } + \frac{n}{\log^4 \kappa} \left( \frac{\ri}{96}+\frac{5 \ri n}{128}+\frac{ H_{n-1}}{48 \pi }+\frac{5 n H_{n-1}}{64 \pi } \right) \nonumber\\
& \left. \phantom{{} = {} } -  \frac{5n}{\log^5 \kappa} \left( \frac{\ri}{128}+\frac{ H_{n-1}}{64 \pi }  \right)  \right]\ .
\end{align}
For $n=1$, the above expression specializes to
\begin{align}\label{genus2kappa}
g_{s}^{-1}\langle W_{n=1}^{1/6}\rangle_{g=2} & =-\frac{\ri \kappa k}{2 \pi }\left[
		-\frac{7  \log \kappa }{720 \pi } + \frac{7 \ri }{1440}-\frac{1}{720 \pi }
		+\frac{1}{\log \kappa } \left( -\frac{\ri}{72} +\frac{1}{36 \pi } \right)
		+\frac{1}{\log^2\kappa} \left( \frac{7 \ri}{288}-\frac{7}{192 \pi } \right) \right. \nonumber\\
& \phantom{= {}}	+	\frac{1}{\log^3\kappa}\left(-\frac{43 \ri}{1152} +\frac{5}{192 \pi }\right)
	+\frac{19 \ri }{384 \log^4\kappa}
	\left.-\frac{5 \ri }{128 \log^5\kappa}
	\right]\ .
\end{align}
The 't Hooft expansion at strong coupling at genus two is found from (\ref{airy}),
\begin{eqnarray}
\langle W_{n}^{1/6}\rangle_{g=2}&=&\frac{\ri^{n+1}\lambda}{(2\pi)^{3}}\Bigg(\frac{7\pi^{3}n^{3}}
{45\sqrt{2\lambda}}
-\Big(\frac{\pi^{2}n^{2}(7n H_{n}+15n-23)}{90}+\frac{7\ri\pi^{3}n^{3}}{180}+\frac{7\pi^{4}n^{4}}{2160}\Big)
\frac{1}{\lambda}\nonumber\\
&&+\Big(\frac{\pi n(3n+1)(3n-7)}{72}+\frac{\pi n^{2}(3n+1)H_{n-1}}{18}+\frac{\ri\pi^{2}n^{2}(3n+1)}{36}\\
&&+\frac{\pi^{3}n^{3}(7n H_{n}+15n-30)}{2160}+\frac{7\ri\pi^{4}n^{4}}{4320}
+\frac{7\pi^{5}n^{5}}{103680}\Big)\frac{1}{\lambda\sqrt{2\lambda}}
+\CO(\lambda^{-2})\Bigg)\re^{\pi n\sqrt{2\lambda}}\ .\nonumber
\end{eqnarray}
The result for the 1/2 BPS Wilson loop is immediately obtained by applying (\ref{sumw}) to (\ref{genus2}), as in the previous cases. 

We can now use the results of section \ref{thooftwilson} for the 't Hooft expansion of 1/2 BPS Wilson loops, 
and study the expression derived there for genus two in the strong coupling region. We have checked explicitly that the strong coupling expansion of $W_2(p)$ agrees with the 
vev for the 1/2 BPS Wilson loop obtained from the Fermi gas result (\ref{genus2}). 

Since we have the exact result (up to exponentially small corrections) for the Wilson loop correlator (\ref{airy}), we can extract the leading and next to the leading terms of the 't~Hooft expansion at arbitrary genus and strong coupling, as it was done in \cite{dg} for the 1/2 BPS Wilson loop of $\CN=4$ super Yang--Mills theory.
For the 1/6 BPS Wilson loop correlator with arbitrary winding $n$, we find
\begin{eqnarray}
\langle W_{n}^{1/6}\rangle_{g}&=&-\ri^{n+1} \frac{n^{2g-1}}{\sqrt{2}}\,a_{g}\sqrt{\lambda}\,\re^{\pi n\sqrt{2\lambda}}\nonumber\\
&&+\ri^{n+1}\frac{n^{2g-2}}{2\pi}\Bigg[\Big(n H_{n-1}+\ri\frac{\pi n}{2}+\frac{\pi^{2} n^{2}}{24}\Big)a_{g}+\frac{3n+1}{12}a_{g-1}+c_{g}\Bigg]\frac{\re^{\pi n\sqrt{2\lambda}}}{\sqrt{\lambda}}\nonumber\\
&&+O(\lambda^{-3/2})\re^{\pi n\sqrt{2\lambda}}\ ,
\end{eqnarray}
where $a_{g}$ and $c_{g}$ are given by
\begin{eqnarray}
&& a_{g}=\frac{2(2^{2g-1}-1)}{(2g)!}B_{2g}\ ,\\
&& c_{g}=\sum_{m=0}^{g}\frac{2(2^{2m-1}-1)2^{2(g-m)}}{(2m)!(2g-2m)!}B_{2m}B_{2g-2m}
\ .
\end{eqnarray}
\par
Using (\ref{sumw}), the leading and next to the leading terms of the 1/2 BPS Wilson loop correlator are found
\begin{eqnarray}
\langle W_{n}^{1/2}\rangle_{g}&=&-\frac{n^{2g-1}}{4}a_{g}\,\re^{\pi n\sqrt{2\lambda}}\nonumber\\
&&+\frac{n^{2g-2}}{2\pi}\Big(\frac{\pi^{2}n^{2}}{48}a_{g}+\frac{3n+1}{24}a_{g-1}\Big)\frac{\re^{\pi n\sqrt{2\lambda}}}{\sqrt{2\lambda}}+\CO(\lambda^{-2})\re^{\pi n\sqrt{2\lambda}}\ .
\end{eqnarray}
It turns out that the 1/2 Wilson loop correlator does not involve $c_{g}$ coefficients. At every genus, the ratio of the leading terms of the 1/6 and 1/2 Wilson loop expectation values is given by
\begin{eqnarray}\label{ratio16-12}
\frac{\langle W_{n}^{1/6}\rangle_{g}}{\langle W_{n}^{1/2}\rangle_{g}}=\frac{\ri^{n+1}\sqrt{\lambda}}{4\sqrt{2}}+\CO(\lambda^{0})\ .
\end{eqnarray}
This ratio was first found in \cite{mpwilson} at genus zero for the trivial winding, and (\ref{ratio16-12}) generalizes this result for arbitrary genus and winding.

There are two interesting properties of the expression (\ref{airy}) and (\ref{airy12}) which are worth noticing. First of all, for a given winding number $n$, 
both expressions are {\it singular} when $k$ is a divisor of $2n$. In particular, for any $n$, they are singular for 
\be
k=1,\, 2.
\ee
Notice that, for these values of $k$, the semiclassical expression for the vev (\ref{GCvev}), which is simply an integral over phase space, is not convergent, and our final result
is reflecting this through a pole in the $\csc$ function. We conclude that our expression is not valid for these special values of $k$. Notice as well that the values $k=1,2$ are certainly special, since
precisely for those values ABJM theory has enhanced supersymmetry to $\CN=8$ \cite{halyo}. As we have just tested, the Fermi gas result resums the genus expansion for $k$ large, therefore the value $k=2n$ sets the convergence radius for this expansion, and the resulting function has the singularities described above. It would be very interesting to understand more 
precisely what happens for these special values of $k$.

Second, we recall that the vevs of 1/2 BPS Wilson loops in ABJM theory can be related to open topological string amplitudes in local $\IP^1 \times \IP^1$ \cite{mpwilson,dmp}.
The expression (\ref{airy12}) can be
interpreted as saying that the {\it unnormalized} Wilson loop vev is given by the Fourier transform of an unnormalized disk amplitude at large radius,
\be
\label{ft-open}
\int \rd\mu\, \re^{-\mu N} \Xi(\mu) {\re^{2n \mu \over k} \over 4 \sin {2 \pi n \over k} },
\ee
where $n$ is the multicovering degree.
As pointed out in \cite{mp}, the Fourier transform expression (\ref{four-t}) for the canonical partition function
can be interpreted as the change of symplectic frame from the large radius topological
string partition function, to the orbifold partition function \cite{abk}. The result (\ref{airy12}) indicates
that a similar result should be valid for the open sector, namely, that changes of frame in the open sector
are also implemented by Fourier transforms of the open string partition function.

We can now identify the overall inverse sine appearing in this formula: it is nothing
but the well-known all-genus bubbling factor for a disk in topological string
theory. This was first found in \cite{ov} by using large $N$ duality with Chern--Simons theory and derived in \cite{kl} with localization techniques. Amusingly, we have
re-derived this factor here by using Sommerfeld's expansion for Fermi gases at low temperature. This bubbling factor resums the genus expansion, as in the Gopakumar--Vafa representation of the
closed topological string free energy \cite{gv}. We see however that this resummation leads to singularities when $k$ is a divisor of $2n$, and this was already observed in the closed string sector in an attempt to resum worldsheet instantons in ABJM theory {\it \`a la} Gopakumar--Vafa \cite{mpu}.

\sectiono{Conclusions and prospects for future work}

In this paper we have used the Fermi gas approach developed in \cite{mp} to compute vevs of Wilson loop observables in the ABJM matrix model of \cite{kwy}. This approach is
based on reformulating the matrix integral as the partition function of an ideal Fermi gas in an exterior potential, and on a semiclassical evaluation of the resulting
quantities. The calculation of Wilson loop vevs is, however, more difficult than the one of the canonical partition function made in \cite{mp}, since one has to resum an infinite number
of corrections in $\hbar$. We have seen however that such a resummation is feasible and we have obtained expressions valid at all orders in the genus expansion and for strong
't Hooft coupling. Equivalently, the expressions we have obtained are full M-theoretic, since they are valid for finite $k$ and large $N$.

Clearly, this work can be generalized in many different ways. Already in ABJM theory, it would be interesting to extend our calculation to higher representations. The Wilson loop operator for a representation with $\ell$ boxes is an $\ell$-body operator in the
Fermi gas. Let us consider for example a 1/6 BPS Wilson loop in the antisymmetric representation. We can write the corresponding matrix model operator as
\be
\CO_{\tableau{1 1}}=\sum_{i<j} \re^{\lambda_i + \lambda_j} ={1\over 2} \sum_{i \not=j} \re^{\lambda_i + \lambda_j}
\ee
so it is clearly a two-body operator. We then have, schematically,
\be
\ba
\langle \CO_{\tableau{1 1}} \rangle^{\rm GC}&={1\over 2} \int \rd \lambda \rd \lambda' \, \re^{\lambda+\lambda'} \rho_2^{\rm GC}(\lambda,\lambda')\\
&={1\over 2} \int \rd \lambda \rd \lambda' \,\left[ n(\lambda, \lambda) n(\lambda',\lambda') -n(\lambda,\lambda') n(\lambda',\lambda)\right]\re^{\lambda+\lambda'}
\ea
\ee
which is a sum of ``direct'' (Hartree) and ``exchange'' (Fock) terms. The first term factorizes into the product of two one-body operator vevs, like the ones computed in this paper,
and we are left with the calculation of the exchange term. This can in principle be computed as well with semi-classical techniques (see \cite{mp-int} for a closely related example), therefore it would
be interesting to develop these techniques further.

Another obvious generalization of this work would be to consider Wilson loop vevs in the $\CN=3$ Chern--Simons--matter theories analyzed from the Fermi gas point of
view in \cite{mp}. In these theories it is general difficult to obtain results by using the traditional tools of matrix models in the 't Hooft expansion, therefore the procedure developed in this paper
is probably the simplest one to go beyond the large $N$ limit. However, a detailed analysis will require computing quantum corrections to the Hamiltonian, Wigner--Kirkwood corrections, and finding an
appropriate regularization of the resummed semiclassical expansion. In fact, it would be interesting to understand in more detail the regularization procedure developed in section 4. In ABJM, 
this procedure could be checked against the results in the 't Hooft expansion, but for more complicated $\CN=3$ theories we might need a better understanding of this issue.

Our final result for 1/2 BPS Wilson loops in terms of Airy functions(\ref{airy12}), gives an analytic result for finite $k\not= 1,2$ and large $N$, at all orders in $1/N$, and up to exponentially small corrections.
Such an expression is very well suited for the type of numerical testing performed in \cite{hanada}. In that paper, numerical calculations provided a beautiful verification of the analytic formulae proposed in \cite{dmp,fhm,mp}, and helped in clarifying certain aspects of the analytic results (like for example the nature of the function $A(k)$ introduced in \cite{mp}). Such a numerical test would be also very useful in understanding what happens when $k=1,2$, where our formula displays a singular behavior.

Finally, we pointed out that (\ref{airy12}) has a natural interpretation as a generalization of the Fourier transform of \cite{abk} to the open sector. It would be clearly very interesting to understand this better, and more generally, to develop techniques to compute topological open string amplitudes at higher genus in a more efficient way.

\section*{Acknowledgements}
A.K. thanks Min-xin Huang and Stefan Theisen for fruitful discussions.
M.M. would like to thank Fernando Casas and Shotaro Shiba for useful communications. A.K.,
M.Sch. and M.S. are supportted by the DFG grant KL2271/1-1. The work of M.M. is supported by the Fonds National Suisse,
subsidies 200020-126817 and 200020-137523.

\appendix

\sectiono{$1/6$ BPS Wilson loops at arbitrary winding number}

In this appendix, we present the details of the matrix model computation which led to (\ref{leading-in}). Our starting point is the integral (\ref{in-int}). This integral can be explicitly evaluated in terms of elliptic functions \cite{handbook}
\begin{equation}
\CI_n=\frac{1 }{2}  (- b)^n \frac{2 \sqrt{a b}}{1+ a b} \sum_{j=0}^n \left( - \frac{a + b}{b} \right)^j \binom {n}{j}V_j\ ,
\end{equation}
where the functions $V_j$ are defined recursively, in terms of elliptic integrals, as follows
\be
\ba
 V_0 &= K(k),\\
 V_1 &= \Pi(\alpha^2, k), \\
 V_2 &= \frac{1}{2(\alpha^2 - 1)(k^2 - \alpha^2)} \left[ \alpha^2 E(k) + (k^2 - \alpha^2)K(k) + (2 \alpha^2 k^2 + 2 \alpha^2 - \alpha^4 - 3 k^2)  \Pi(\alpha^2, k) \right],\\
 V_3 & = \frac{1}{ 2(m+2)(1- \alpha^2)(k^2 - \alpha^2) } \left[ (2 m + 1)k^2 V_m + (2 m + 1)(\alpha^2 k^2 + \alpha^2 - 3 k^2)V_{m+1}   \right. \\
	& \left. + (2 m +3)(\alpha^4 - 2 \alpha^2 k^2 - 2 \alpha^2 + 3 k^2)V_{m+2} \right].
\ea
\ee
In the above expressions, the moduli of the elliptic functions are defined by
\begin{equation}
 k^2 = \frac{(a^2 - 1)(b^2 -1)}{ (1 + ab)^2}\ , \qquad \alpha^2 = \frac{1 - a^2}{ 1 + ab}\ .
\end{equation}
To understand the strong coupling behavior of the integral (\ref{in-int}), we need to expand the above functions in the large $\kappa$ regime. First, we notice that
\be
\ba
 V_0 & \approx 2 \log \kappa, \\
 V_1 & \approx \frac{\kappa}{16}(\pi - 6 \ri \log \kappa), \\
 V_2 & \approx \frac{\kappa^2}{32}\left( 1 - \frac{\ri \pi}{2} - 3 \log \kappa \right),
 \ea
 \ee
and the recursion relation becomes, at large $\kappa$,
\be
V_{3+m} \approx \frac{1 + m}{16(2+m)}\kappa^2 V_{m+1} - \frac{\ri(3 + 2m)}{4(2+m)} \kappa V_{m+2}\ .
\ee
We can easily find the solution to the above recursion relation. We have
 \begin{equation}\label{solrecur}
 V_j \approx \left( \frac{\kappa}{4 \ri} \right)^j \left( \frac{ H_{j-1} }{2} + \frac{\ri \pi}{4} + \frac{3}{2}\log \kappa \right), \qquad j \geq 1 \,.
\end{equation}
Using the above solution (\ref{solrecur}), we then compute the integral (\ref{in-int}) in regime of large $\kappa$
\begin{equation}\label{In-sum}
\CI_n\approx \ri^n  \kappa^{n-1} \left( \sum_{j=1}^n \left(  - \frac{ H_{j-1} }{2} + \frac{\ri \pi}{4}
 + \frac{3}{2}\log \kappa \right) (-1)^j \binom{n}{j} + 2 \log \kappa \right)\ .
\end{equation}
In order to perform the sum on the harmonic numbers in the above expression, we use the following integral representation of harmonic numbers
\begin{equation}
 H_{j-1} = \int_0^1 \frac{1 - x^{j-1}}{1 - x} \rd x\ .
\end{equation}
Using the above representation, we then obtain
\begin{equation}
\sum_{j=1}^n  (-1)^j \binom{n}{j}H_{j-1 } = \int_0^1 \frac{\rd x}{x} \left( 1 - (1-x)^{n-1} \right) =H_{n-1}\ .
\end{equation}
The sum on the rest of the terms in (\ref{In-sum}) is easy to perform
\begin{equation}
\sum_{j=1}^n (-1)^j \begin{pmatrix} n \\ j \end{pmatrix} \left( \frac{\ri \pi}{4} + \frac{3}{2}\log \kappa \right)  + 2 \log \kappa = - \frac{\ri \pi}{4} + \frac{1}{2}\log \kappa\ .
\end{equation}
Putting things together, we therefore conclude
\begin{equation}
 \CI_n \approx \ri^n  \kappa^{n-1}  \left(  -  \frac{H_{n-1}}{2} -\frac{ \ri \pi}{4} + \frac{1}{2}\log \kappa  \right),
\end{equation}
which is the sought-for result (\ref{leading-in}).

\sectiono{Results at $g=3$ and $g=4$}
It is evident from (\ref{airy}) that extracting higher genus contributions to the expectation values of the ABJM Wilson loops is more economical than other existing approaches, such as the matrix model approach. To demonstrate this, we summarize the result of the 't~Hooft expansion for $g=3$ and $g=4$ in this appendix. It will not be necessary to find the expansions in terms of $\kappa$, as there is no matrix model computation that we can compare against it. It will be sufficient to directly expand the Airy functions in (\ref{airy}) in terms of the 't~Hooft coupling at strong coupling regime.
\par
For $g=3$, we obtain
\begin{eqnarray}\label{genus3}
\langle W_{n}^{1/6}\rangle_{g=3}&=&-\frac{\ri^{n+1}\lambda}{(2\pi)^{5}}\Bigg(\frac{62\pi^{5}n^{5}}
{945\sqrt{2\lambda}}-\Big(\frac{\pi^{4}n^{4}(62nH_{n}+147n-323)}{1890}+
\frac{31\ri\pi^{5}n^{5}}{1890}+\frac{31\ri\pi^{6}n^{6}}{22680}\Big)\frac{1}{\lambda}
\nonumber\\
&& +\Big(\frac{\pi^{3}n^{3}(3n+1)(14n H_{n}+15n-65)}{540}+\frac{7\ri\pi^{4}n^{4}(3n+1)}{540}
\\
&&+\frac{\pi^{5}n^{5}(62nH_{n}+147n-385)}{45360}
 +\frac{31\ri\pi^{6}n^{6}}{45360}+\frac{31\pi^{7}n^{7}}{1088640}\Big)
\frac{1}{\lambda\sqrt{2\lambda}}+\CO(\lambda^{-2})\Bigg)\re^{\pi n\sqrt{2\lambda}}\ ,\nonumber
\end{eqnarray}
while the $g=4$ contribution is given by
\begin{eqnarray}\label{genus4}
\langle W_{n}^{1/6}\rangle_{g=4}&=&\frac{\ri^{n+1}}{(2\pi)^{7}}\Bigg(\frac{127\pi^{7}n^{7}}
{4725\sqrt{2\lambda}}-\Big(
\frac{\pi^{6}n^{6}(381n H_{n}+930n-2738)}{28350}+\frac{127\ri\pi^{7}n^{7}}{18900}+\frac{127\pi^{8}n^{8}}{226800}
\Big)\frac{1}{\lambda}\nonumber\\
&& +\Big(\frac{\pi^{5}n^{5}(3n+1)(124n H_{n}+147n-819)}{11340}+\frac{31\ri\pi^{6}n^{6}(3n+1)}{5670}\\
&&+\frac{\pi^{7}n^{7}(381n H_{n}+930n-3119)}{680400}
+\frac{127\ri\pi^{8}n^{8}}{453600}+\frac{127\pi^{9}n^{9}}{10886400}
\Big)\frac{1}{\lambda\sqrt{2\lambda}}+\CO(\lambda^{-2})\Bigg)\re^{\pi n\sqrt{2\lambda}}\ .\nonumber
\end{eqnarray}
Applying (\ref{sumw}), we can immediately find the result for the 1/2 BPS Wilson loop correlator at $g=3$ and $g=4$.

\end{document}